\documentclass[twocolumn,prd,showpacs,preprintnumbers,amsmath,amssymb,superscriptaddress,floatfix,altaffilletter,nofootinbib]{revtex4-1}
\usepackage{graphicx}
\usepackage{multirow}
\usepackage{subfigure}
\usepackage{color}
\definecolor{MyGrey}{rgb}{0,0,0} 
\definecolor{MyDarkBlue}{rgb}{0.23,0.21,0.69} 
\definecolor{MyLightBlue}{rgb}{0.22,0.51,0.86}
\usepackage[colorlinks=true,linkcolor=MyGrey,citecolor=MyGrey,urlcolor=MyGrey,bookmarksnumbered=true,bookmarksopen]{hyperref}
\newcommand{\be}{\begin{equation}}
\newcommand{\ee}{\end{equation}}
\newcommand{\bea}{\begin{eqnarray}}
\newcommand{\eea}{\end{eqnarray}}
\begin{document}
\begin{titlepage}
\title{Numerical implementation of lepton-nucleus interactions\\ and its effect on neutrino oscillation analysis}
\author{C.-M.~Jen}\affiliation{Center for Neutrino Physics, Virginia Tech, Blacksburg, VA 24061, USA}
\author{A.~M.~Ankowski}
\altaffiliation[Now at ]{Center for Neutrino Physics, Virginia Tech, Blacksburg, VA 24061, USA}
\affiliation{Department of Physics, Okayama University, Okayama 700-8530, Japan}
\author{O.~Benhar}
\altaffiliation[On leave from ]{INFN and Department of Physics, ``Sapienza'' Universit\`a di Roma, I-00185 Roma, Italy.}
\affiliation{Center for Neutrino Physics, Virginia Tech, Blacksburg, VA 24061, USA}
\author{A.~P.~Furmanski}
\affiliation{University of Warwick, Department of Physics, Coventry, United Kingdom}
\author{L.~N.~Kalousis}
\affiliation{Center for Neutrino Physics, Virginia Tech, Blacksburg, VA 24061, USA}
\author{C.~Mariani}
\affiliation{Center for Neutrino Physics, Virginia Tech, Blacksburg, VA 24061, USA}
\date{\today}
\begin{abstract}
We discuss the implementation of the nuclear model based on realistic nuclear spectral functions in the GENIE neutrino interaction generator. Besides improving on the Fermi gas description
of the nuclear ground state, our scheme involves a new prescription for $Q^2$ selection, meant to efficiently enforce energy-momentum conservation.
The results of our simulations, validated through comparison to electron scattering data, have been obtained for a variety of target nuclei, ranging from carbon to argon, and cover the kinematical region in which quasielastic scattering is the dominant reaction mechanism. We also analyze the influence of the adopted nuclear model on the determination of neutrino oscillation parameters.
\end{abstract}
\maketitle
\end{titlepage}

\section{Introduction}
\label{sec:intro}
Neutrino physics is entering the age of precision measurements.  Several experiments have detected
neutrino oscillations, providing unambiguous evidence that neutrinos---assumed to be massless in the standard model of particle
physics---have nonvanishing masses.
The recent observations of a large $\theta_{13}$ mixing angle, reported by the Double Chooz~\cite{Abe:2011fz}, Daya Bay~\cite{An:2013zwz}, RENO~\cite{Ahn:2012nd}, and T2K~\cite{Abe:2013hdq} Collaborations,
entail the possibility of measuring CP violation in the leptonic sector, thus addressing one of the outstanding
problems of particle physics.  However, these measurements will involve high precision determinations of the oscillation
parameters, which in turn require a deep understanding of neutrino interactions with matter.
In view of the achieved and expected experimental accuracies, the treatment of nuclear effects is in fact one of the
main sources of systematic uncertainty~\cite{Abe:2013xua}.

Over the past decade, several experiments \cite{miniboone_ccqe,miniboone_nc,K2K_ccqe} have unambiguously exposed the inadequacy
of the relativistic Fermi gas model (RFGM), routinely employed in simulation codes of neutrino interactions, to reproduce the observed cross sections.
As a consequence, a great deal of effort has
been devoted to the development of more realistic descriptions of nuclear effects~\cite{PhysRevD.72.053005,Benhar:2006nr,Martini:2012fa,Meloni:2012fq,Nieves:2012yz,Lalakulich:2012hs, Martini:2012uc, Mosel:2013fxa}.
In this context, a pivotal role is played by the availability of a large body of
theoretical and experimental studies of electron-nucleus scattering.

Accurate measurements of the coincidence $(e,e^\prime p)$ cross section have provided
quantitative information on nuclear spectral functions, revealing the limitations of the
independent particle model of the  nucleus. While the spectroscopic lines corresponding to knock out of nucleons in shell
model states are in fact clearly visible in the missing energy spectra, the associated spectroscopic factors are
considerably lower than expected, regardless of the nuclear mass number. This is a clear manifestation of the importance of
correlations, that lead to the excitation of nucleon-nucleon pairs to states of energy larger than the Fermi energy, thus depleting
the shell-model states within the Fermi sea. Comparison between the results of theoretical calculations and electron scattering data have
provided overwhelming evidence that correlation effects~\cite{RevModPhys.65.817,RevModPhys.69.981} must be included in any realistic descriptions of nuclear interactions.

The extension of the theoretical description of electron-nucleus scattering to the case of
neutrino interactions does not involve severe conceptual difficulties.
However, while significant progress has been made in the understanding of the
different reaction mechanisms contributing to the signals detected by neutrino experiments, the
implementation of state-of-the-art models in the existing Monte Carlo generators has been lagging behind.

The first step towards an improved treatment of nuclear effects is the replacement of the
RFGM with a more realistic description of the nuclear ground state, based on spectral functions
obtained from advanced many-body approaches. It has to be emphasized that a better modeling
of the {\em initial} state is of paramount importance, as it obviously affects {\em all} reaction channels.

In this article, we discuss the implementation of the nuclear spectral functions of Refs.~\cite{LDA,PhysRevD.72.053005,Ankowski:2007uy} in the GENIE
neutrino interaction generator. We also analyze the significance of the description of the nuclear ground state for
the determination of the oscillation parameters. Our study is focused on the charged-current quasielastic (CCQE)
channel, which accounts for a large fraction of the detected signal in many experiments.

In Section~\ref{sec:theory} we outline the elements of the calculation of the electron- and neutrino-nucleus
cross section in the kinematical regime in which the impulse approximation is expected to be applicable.
The implementation of the nuclear model based on spectral functions into
the GENIE event generator, as well as its validation through comparison to electron-nucleus  scattering data
are discussed in Sec.~\ref{subgenie}. Section~\ref{sec:oscillation} is devoted to the analysis of the
impact of the description of nuclear dynamics on the determination of the neutrino oscillation parameters. Finally,
in Sec.~\ref{sec:conclusions} we summarize the main results of our work and state the conclusions.

\vspace{0.3cm}

\section{Quasielastic Electron- and Neutrino-Nucleus Cross Sections}
\label{sec:theory}

\vspace{0.2cm}

This Section is devoted to the description of our numerical implementation of the lepton-nucleus interaction model discussed in Refs.~\cite{Benhar:2006wy, PhysRevD.72.053005}.

The procedure employed to obtain the cross sections involves all the elements required to carry out a simulation of the scattering process.
Therefore, our results can be used as benchmarks, to test the predictions of any event generators  based on the same dynamical model and describing the same reaction mechanisms.
\par Within the impulse approximation (IA), which is expected to be applicable at momentum transfer $|\bf q|$ such that $1/|{\bf q}| \lesssim d$, $d$ being the average nucleon-nucleon separation distance, nuclear scattering reduces to the incoherent sum of elementary scattering processes involving individual particles. As a first approximation, the anti-symmetrization of the final nuclear state and the occurrence of final-state interactions (FSI) between the nucleon interacting with the beam particle and the spectator nucleons will be neglected. These effects, as well as more complex mechanisms not included in the IA picture, will not be analyzed in this article.

The building blocks of the calculation discussed here are:
\begin{itemize}
\item [{(a)}] The description of the {\em initial state}, based on a model of nuclear dynamics. Initial state dynamics determines the target spectral function, yielding the energy and momentum distribution of the target nucleons.
\item  [{(b)}] The description of the {\em elementary interaction vertex}. For any given values of the beam energy and nucleon four-momentum, the interaction vertex determines the kinematical variables associated with the outgoing particles.
\end{itemize}
%
\begin{figure}
\subfigure[]{\includegraphics[width=1.0\columnwidth]{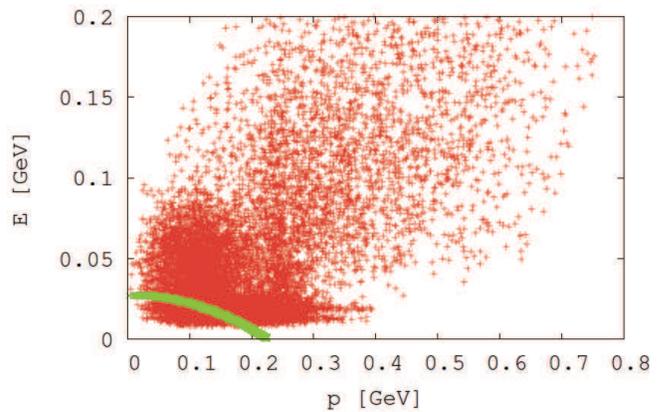}\label{fig:F00a}}
\subfigure[]{\includegraphics[width=1.0\columnwidth]{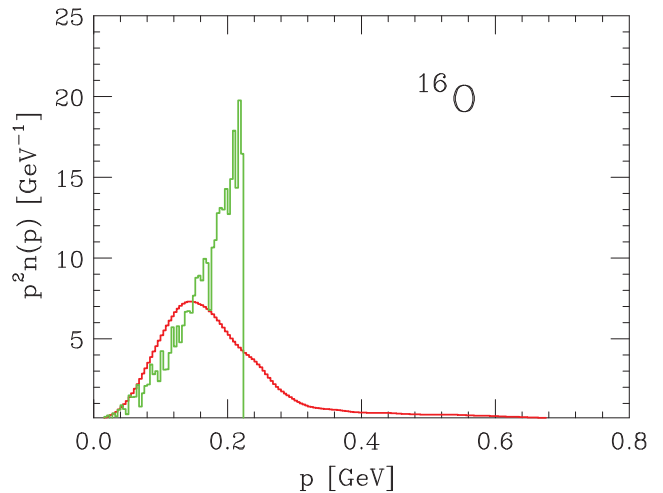}\label{fig:F00b}}
\caption{(Color online). (a): distribution of  20,000 $(p,E)$ pairs sampled from the probability distribution \eqref{prob:dist} using the oxygen spectral
function of Ref.~\cite{ PhysRevD.72.053005} (red) and the RFGM with $p_F=209$~MeV and $\epsilon_0=27$~MeV (green).
(b): probability distribution of nucleon momentum $p^2 n(p)$,  defined as in Eq.~\eqref{def:np}, obtained from the spectral of function of Ref.~\cite{ PhysRevD.72.053005} (red) and the RFGM (green).\label{F00}}
\end{figure}
%
\subsection{Initial State}\label{IS}
The initial state of the target is described by the spectral function $P({\bf p},E)$,  yielding the probability of removing a nucleon of momentum ${\bf p}$ from the target nucleus, leaving the residual system with excitation energy $E$. From this definition, it follows that the energy of the residual $(A-1)$-nucleon system can be written in the form
\begin{equation}
\label{E:R}
E_{A-1} = \sqrt{ (M_A - m + E)^2 + p^2 } \ ,
\end{equation}
where $p= |{\bf p}|$ and $M_A$ and $m$ are the target and nucleon mass, respectively. Note that, owing to nucleon-nucleon correlations, the state of the residual system is {\em not} restricted to be a bound state.
\par Figure~\ref{fig:F00a} shows the distribution of 20,000 $(p,E)$ pairs, obtained sampling the function
\begin{equation}
\label{prob:dist}
F(p,E) = 4 \pi p^2 P(p,E)  \ ,
\end{equation}
using the oxygen spectral function of Ref.~\cite{ PhysRevD.72.053005}, constructed combining $(e,e^\prime p)$ data and {\em ab initio}
nuclear matter calculations within the local-density approximation (LDA)~\cite{LDA}.
It clearly appears that it extends well beyond the region of the $(p,E)$ plane
spanned by the shell-model predictions.
\par
Within the RFGM, the spectral function is parametrized in the simple form
\begin{equation}
\label{PkE:FG}
P(p,E) = \frac{3}{4 \pi p^3_F}  \ \theta(p_F - p) \ \delta( E + \sqrt{ p^2 + m^2 } -m  - \epsilon_0 )  \ ,
\end{equation}
$p_F \sim 209 \ {\rm MeV}$ and $\epsilon_0 \sim 27 \ {\rm MeV}$ being the Fermi momentum and the average nucleon binding energy, respectively, and the distribution of Fig.~\ref{fig:F00a} collapses to a line (the spread visible in the figure arises form the finite width of the energy and momentum bins).
\par In Fig.~\ref{fig:F00b}, the probability distribution of the nucleon momentum
\begin{equation}
\label{def:np}
p^2 n(p) = p^2 \int dp^\prime  dE  \ \delta(p - p^\prime) \  F(|{\bf p}^\prime|,E)  \ ,
\end{equation}
obtained from the 20,000 $(|{\bf p}|,E)$ samples of Fig.~\ref{fig:F00a} is compared to the RFGM prediction corresponding to $p_F=209$ MeV.

\par
Note that the available spectral functions depend on the {\em magnitude} of the nucleon momentum only. Taking into account the angular dependence of the momentum distribution of non spherical nuclei (e.g. $^{12}$C) involves considerable difficulties, mainly arising from the correlation between polar angle and nucleon energy.

The LDA spectral functions~\cite{LDA,PhysRevD.72.053005} are available for carbon, oxygen, and iron.
Unfortunately, in the case of calcium $^{40}_{20}$Ca the information provided by $(e,e'p)$ measurements is scarce,
and for argon $^{40}_{18}$Ar there is no information at all. As a consequence, the LDA procedure cannot be currently applied to those nuclei. The available spectral functions (GSF)~\cite{Ankowski:2007uy} have been obtained from models involving rather crude approximations, and exhibit an oversimplified momentum and energy dependence. A detailed comparison between the oxygen cross sections obtained using LDA~\cite{LDA,PhysRevD.72.053005} and GSF~\cite{Ankowski:2007uy} spectral functions  can be found
in Fig.~5 of Ref.~\cite{Ankowski:2007uy}.
\subsection{Interaction Vertex}\label{V}
The interaction vertex is described by the cross section of the elementary process, involving a {\em bound} {\em moving} nucleon.
It can be written in the general form
\begin{equation}
\label{def:vertex}
\left( \frac{ d^2 \sigma} {d\omega d \Omega_{k^\prime}} \right)_{N} \propto L_{\mu \nu} (k,k^\prime)
W^{\mu \nu}({\widetilde p},  {\widetilde p}+ {\widetilde q}) \ ,
\end{equation}
%
with $q = k - k^\prime \equiv (\omega,{\bf q})$ being the four-momentum transfer. In the case of neutrino [electron] scattering, we denote the four-momenta of the incoming and outgoing lepton as $k \equiv (E_\nu,{\bf k})$ and $k^\prime \equiv (E_\ell,{\bf k}^\prime)$ [$k \equiv (E_e,{\bf k})$ and $k^\prime \equiv (E_{e'},{\bf k}^\prime)$], respectively.
\par
The tensor $L_{\mu \nu}$ depends on lepton kinematical variables only. Its expression for electron scattering reads
\begin{equation}
L_{\mu\nu} = 2\,\left[k_\mu\,k^{'}_\nu+k_\nu\,k^{'}_\mu- g_{\mu\nu}(k \cdot k^{'}) \right] \ ,
\label{eq:tensor_electron}
\end{equation}
with $g_{\mu \nu} = \rm{diag}(1, -1, -1, -1)$, while in the case of charged current neutrino interactions it is given by
\begin{equation}
L_{\mu\nu}=4\,\left[k_\mu\,k^{'}_\nu+k_\nu\,k^{'}_\mu- g_{\mu\nu}(k \cdot k^{'})-i\,\varepsilon_{\mu\nu\alpha\beta}\,k^{'\beta}\,k^\alpha \right]  \ ,
\label{eq:tensor_neutrino}
\end{equation}
where $\epsilon_{\mu\nu\alpha\beta}$ is the fully antisymmetric Levi-Civita tensor.
\par
The tensor $W^{\mu \nu}$ contains all the information on the structure of the target nucleon. In the quasielastic sector its expression involves the nucleon vector and axial-vector form factors.
\par
In principle, $W^{\mu \nu}$ depends on the nucleon initial and final four-momenta
$p \equiv (p_0, {\bf p})$, with $p_0 = M_A - E_{A-1}$ and $p^\prime = p + q$.
It is very important to realize, however, that in lepton-nucleus scattering a fraction
of the energy transfer to the target goes into the excitation energy of the spectator particles. As a consequence,
the energy transfer involved in the elementary interaction can be conveniently written in the
form~\cite{Benhar:2006wy}
\begin{equation}
\label{delta:omega}
{\widetilde \omega } = \omega - \delta \omega \ ,
\end{equation}
where ${\widetilde \omega}$ is the amount of energy required for elastic scattering off a nucleon
carrying momentum ${\bf p}$ {\em in free space}, i.e.
\begin{equation}
\label{omega:tilde}
{\widetilde \omega } = \sqrt{|{\bf p} + {\bf q}|^2 + m^2} - \sqrt{p^2 + m^2} \ .
\end{equation}
Combining the above equation with energy conservation, implying
\begin{equation}
M_A + \omega = \sqrt{|{\bf p} + {\bf q}|^2 + m^2} + E_{A-1} \ ,
\end{equation}
we obtain
\begin{equation}
\label{omega:tilde2}
{\widetilde \omega } = \omega + M_A - E_{A-1} - \sqrt{p^2 + m^2}  \ .
\end{equation}
Note that the physical interpretation of ${\widetilde \omega }$ becomes very transparent in the
$(p/m) \to 0$ limit, yielding ${\widetilde \omega} = \omega - E$.

The introduction of the new variable $\widetilde \omega$ in Eq.~\eqref{def:vertex}, while being fully justified on physics grounds, leads to a violation of gauge
invariance, which is a direct consequence of the assumptions implied in the impulse approximation.
In our calculations, gauge invariance has been restored using the CC1 prescription developed in Ref.~\cite{Forest83}, widely employed in the analysis of $(e,e^\prime p)$ experiments.
Note that, owing to the replacement $\omega \rightarrow \widetilde \omega$ in the nucleon tensor of Eq.~\eqref{def:vertex}, the four-momentum squared
transferred at the interaction vertex explicitly depends on the initial nucleon momentum and energy, since
%
\begin{equation}
{\widetilde p} \equiv( \sqrt{|{\bf p}|^2 + m^2} , {\bf p}) \ \ \ , \ \ \
{\widetilde q} \equiv ( {\widetilde \omega}, {\bf q}) \ ,
\label{eq:omega_tilde}
\end{equation}

It has to be emphasized that expressing the nucleon tensor $W^{\mu \nu}$ as a function of the variables
${\widetilde p}$ and ${\widetilde q}$ allows one to consistently use nucleon structure functions obtained from the {\em measured} proton and deuteron cross sections.

As pointed out above, for any $E_\nu$ (or $E_e$),  ${\bf p}$ and $E$ the elementary cross section is a function of two
variables, e.g. $q = |{\bf q}|$ and ${\omega}$, yielding the probability distribution of the kinematical variables of the outgoing particles.

\begin{figure*}
\subfigure[~$e+^{16}_{~8}$O, $E_{e}=0.88$~GeV, $\theta_{e'}=32^{\circ}$]{\includegraphics[width=0.9\columnwidth]{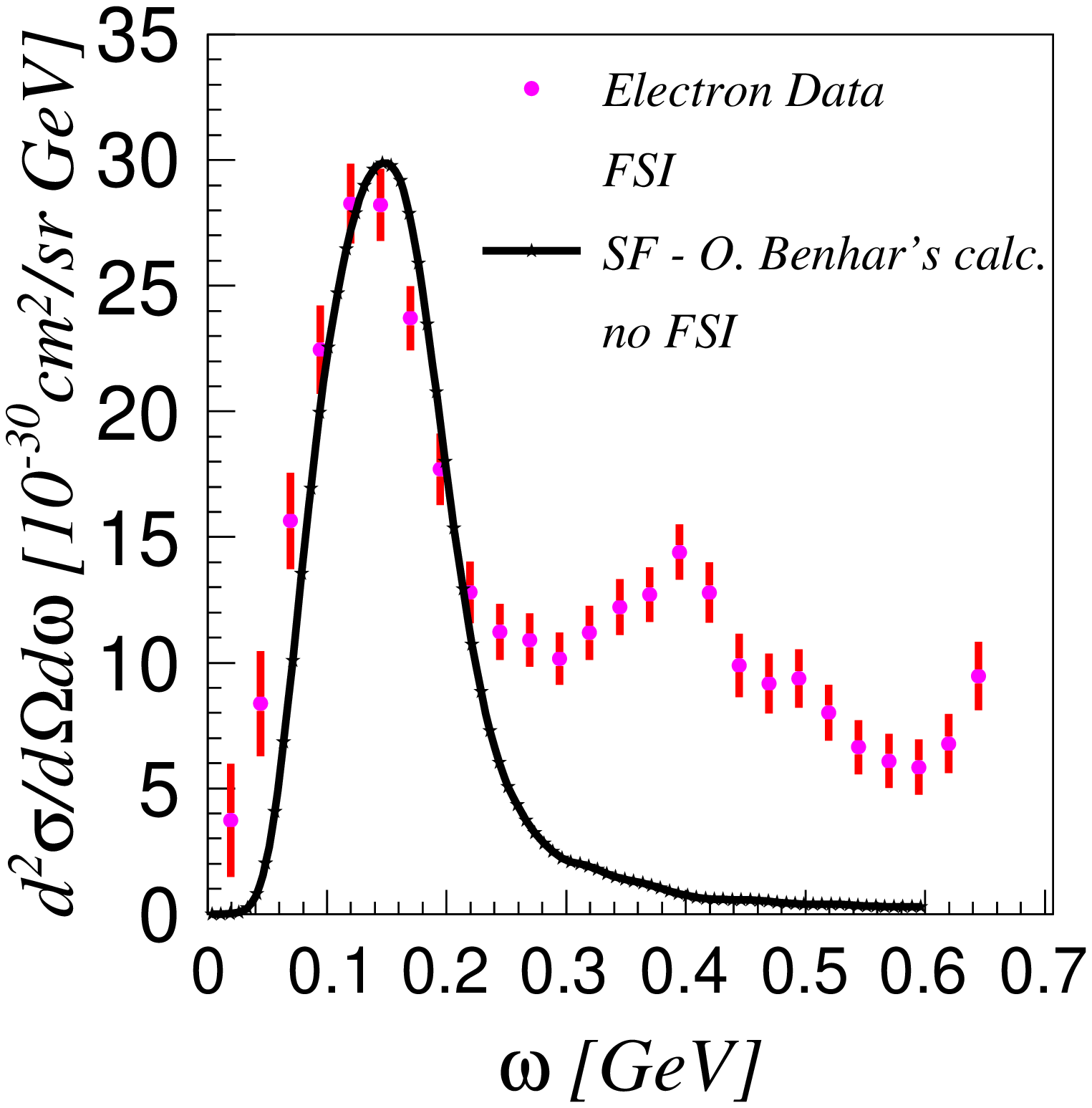}}
\subfigure[~$e+^{16}_{~8}$O, $E_{e}=1.20$~GeV, $\theta_{e'}=32^{\circ}$]{\includegraphics[width=0.9\columnwidth]{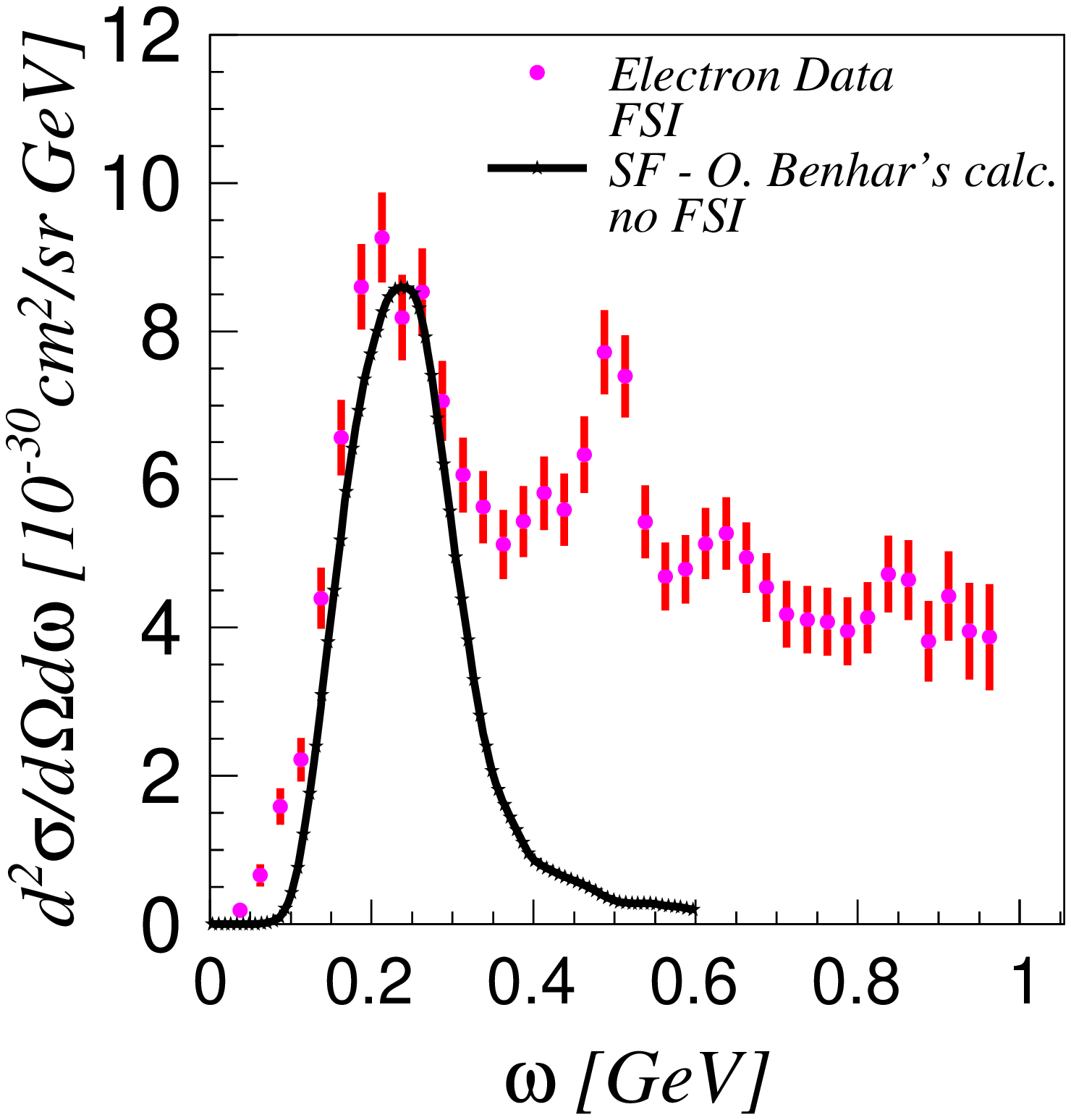}}
\caption{(Color online). Double differential cross section of the process
    $e+{^{16}}\textrm{O} \to e^\prime + X$
    in the quasielastic channel. The calculations have been carried
    out using Eqs.~\eqref{MC1}--\eqref{MC3} with 20,000 $(p,E)$
    pairs sampled from the probability distribution of Eq.~\eqref{prob:dist}
   and  the spectral function of Ref.~\cite{ PhysRevD.72.053005}, without including any modeling of final state interactions.
    The data, taken from Refs.~\cite{Anghinolfi:1996vm,Anghinolfi:95n}
    and available online at {\color{blue} {\protect \url{http://faculty.virginia.edu/qes-archive/index.html}}}, are not corrected to remove the
    effects of FSI.\label{F1}}
\end{figure*}

%
\subsection{Nuclear Cross Section}
The derivation of the double differential nuclear cross section in the IA regime is described in detail in Refs.~\cite{ Benhar:2006wy,PhysRevD.72.053005}. In the quasielastic channel the final result, obtained in the target rest frame, can be cast in the form
\begin{align}
\label{cross:section}
 & \left( \frac{d^2 \sigma}{d\omega d \Omega_{k^\prime}} \right)_A  = \int d^3p dE \left(\frac{d^2\sigma} {d\omega d\Omega_{k^\prime}} \right)_{N} P(|{\bf p}|,E)~ \nonumber \\
& \ \ \ \ \ \ \times~ \delta(\omega + M_A -  \sqrt{ |{\bf p} + {\bf q}|^2 + m^2 } - E_{A-1} )  \ .
\end{align}

The explicit expression of the elementary differential cross section [see Eq.~\eqref{def:vertex}] for electron and charged current neutrino scattering can be found in Refs.~\cite{Benhar:2006wy,Benhar:2006nr}, respectively.

According to the standard representation of electron scattering data, the double differential cross section is given at fixed beam energy
and scattering angle of the outgoing lepton, as a function of energy loss $\omega$.

In order to set a benchmark for the implementation of the spectral function approach into GENIE, we have computed the electron- and neutrino-nucleus cross sections from Eq.~\eqref{cross:section}.

The integration has been carried out using the Monte Carlo approach, yielding
\begin{align}
\label{MC1}
& \left( \frac{d^2 \sigma}{d\omega d \Omega_{k^\prime}} \right)_{A}  = \int dp \ dE \ d \cos \theta_p  \ \ G(k,k^\prime ; p,E, \cos \theta_p)  \nonumber  \\
& \ \ \ \ \ \ \times F(p,E)
\approx  \frac{1}{N} \sum_{n=1}^N G( k,k^\prime ; \{ p ,E , \cos \theta_p \}_n ) \ ,
\end{align}
where $\theta_p$ is the polar angle specifying the direction of the nucleon momentum, ${\bf p}$, and
\begin{align}
\label{MC2}
 & G(k,k^\prime ; p,E, \cos \theta_p)    =  \left( \frac{d^2 \sigma}{d\omega d \Omega_{k^\prime}} \right)_{N} \nonumber \\
& \ \ \ \ \ \ \ \times \delta( \omega + M_A - \sqrt{ |{\bf p} + {\bf q}|^2 + m^2 } - E_{A-1} )  \ .
\end{align}
The above expressions have been evaluated with Monte Carlo configurations $\{ p ,E , \cos \theta_p \}_n$, with $n=1,\ldots ,  N$ and $N=20,000$. The values of $p$ and $E$ have been sampled from the distribution of Eq.~\eqref{prob:dist}, while $\cos \theta_p$ has been sampled from a uniform distribution. The delta function has been implemented using the finite width representation
\begin{equation}
\label{MC3}
\delta (x) =  \frac{1}{2 \sqrt{\pi \epsilon}} \  e^{-x^2/4 \epsilon} \ ,
\end{equation}
providing $\epsilon$-independent results for small $\epsilon$.
\subsection{Electron scattering}\label{electron}
The  form of the lepton tensor  $L_{\mu \nu}$ for electron scattering is given by Eq.~\eqref{eq:tensor_electron}, while
the explicit expression of the nucleon tensor $W^{\mu \nu}$, involving the nucleon vector form factors, can be found in Ref.~\cite{Benhar:2006wy}.

The proton $(p)$ and neutron $(n)$ vector form factors, $F_1^{p,n}$ and
$F_2^{p,n}$, have been precisely measured
up to large values of $Q^2= -q^2$ in electron-proton and electron-deuteron scattering experiments,
respectively (for a recent review see, e.g., Ref.~\cite{Perdrisat:2006hj}).
The results presented in this article have been obtained  using the parametrization referred to as BBBA05~\cite{Bradford:2006yz},
obtained from an analysis including recent measurements carried out at the Thomas Jefferson National Accelerator Facility.

As an example, Fig.~\ref{F1} shows a comparison between the electron-oxygen cross sections computed from Eqs.~\eqref{MC1}--\eqref{MC3} and the experimental data of Refs.~\cite{Anghinolfi:1996vm,Anghinolfi:95n}. It clearly appears that both position and width of the quasielastic bump---dictated by the energy and momentum dependence of the spectral function, respectively---are described with remarkable accuracy. In this respect,  it is worth reminding that the results shown on Fig.~\ref{F1} involve {\em no adjustable parameters} and do not include any modeling of FSI, the effects of which are obviously contained in the data.

\subsection{Neutrino Scattering}\label{neutrino}
The lepton tensor $L_{\mu \nu}$ for charged current neutrino interactions is given by Eq.~\eqref{eq:tensor_neutrino}, while
the expression of the nucleon tensor $W^{\mu \nu}$ can be found in Ref.~\cite{Benhar:2006nr}. In addition to the vector form factors,
in this case the definition of $W^{\mu \nu}$ involves the axial form factor, generally parametrized in the dipole form
\begin{equation}
F_A(Q^2) = \frac{g_A}{(1 + Q^2/M_A^2)^2} \ ,
\label{eq:axial_mass}
\end{equation}
where  $g_A = -1.26$ and the axial mass $M_A$ is the parameter determining the $Q^2$ dependence. Its value, extracted from
(quasi)elastic neutrino and antineutrino-nucleon scattering, charged pion electroproduction off nucleons and muon capture data is
$M_A=1.03$~MeV~\cite{Bernard:2001rs}.

\begin{figure}
\includegraphics[width=\columnwidth]{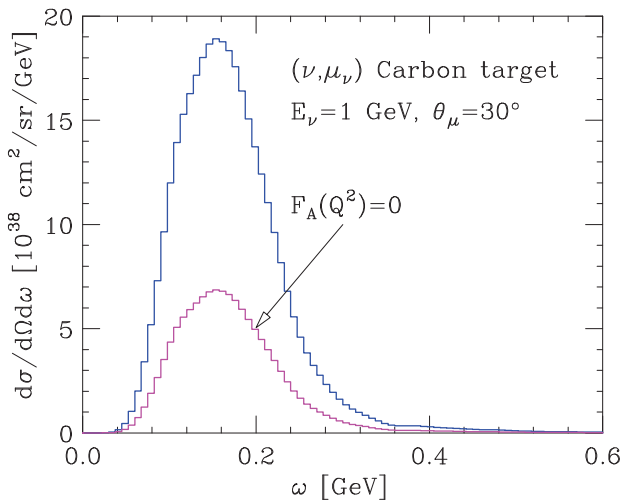}
\caption{(Color online). Double differential cross section of the process $\nu_\mu+{^{12}\textrm{C}} \to \mu^- + X$ in the quasielastic channel, obtained
using the spectral function of Ref.~\cite{LDA}. The two histograms show the results of the full calculation and those obtained setting $F_A(Q^2)=0$.\label{F3}}
\end{figure}

As an example, Fig.~\ref{F3} shows the double differential cross section of the process
\begin{equation}
\nu_\mu + {^{12}\textrm{C}} \to \mu^- + X \ ,
\end{equation}
in the quasielastic channel, at neutrino energy $E_\nu = 1$ GeV and muon scattering angle $\theta_\mu = 30$ deg, plotted as a function of the lepton energy loss $\omega$. The calculation has been carried out  using the carbon spectral function of Ref.~\cite{LDA}. In order to illustrate the size of the axial-vector contributions, the result of the full calculation is compared to that obtained setting $F_A(Q^2)=0$.

%

\begin{figure*}
\begin{center}
\subfigure[~$e+\textrm{C} \to e' + X$, $E_{e}=0.961$~GeV, $\theta_{e}=37.5$~deg]{\includegraphics[width=0.9\columnwidth]{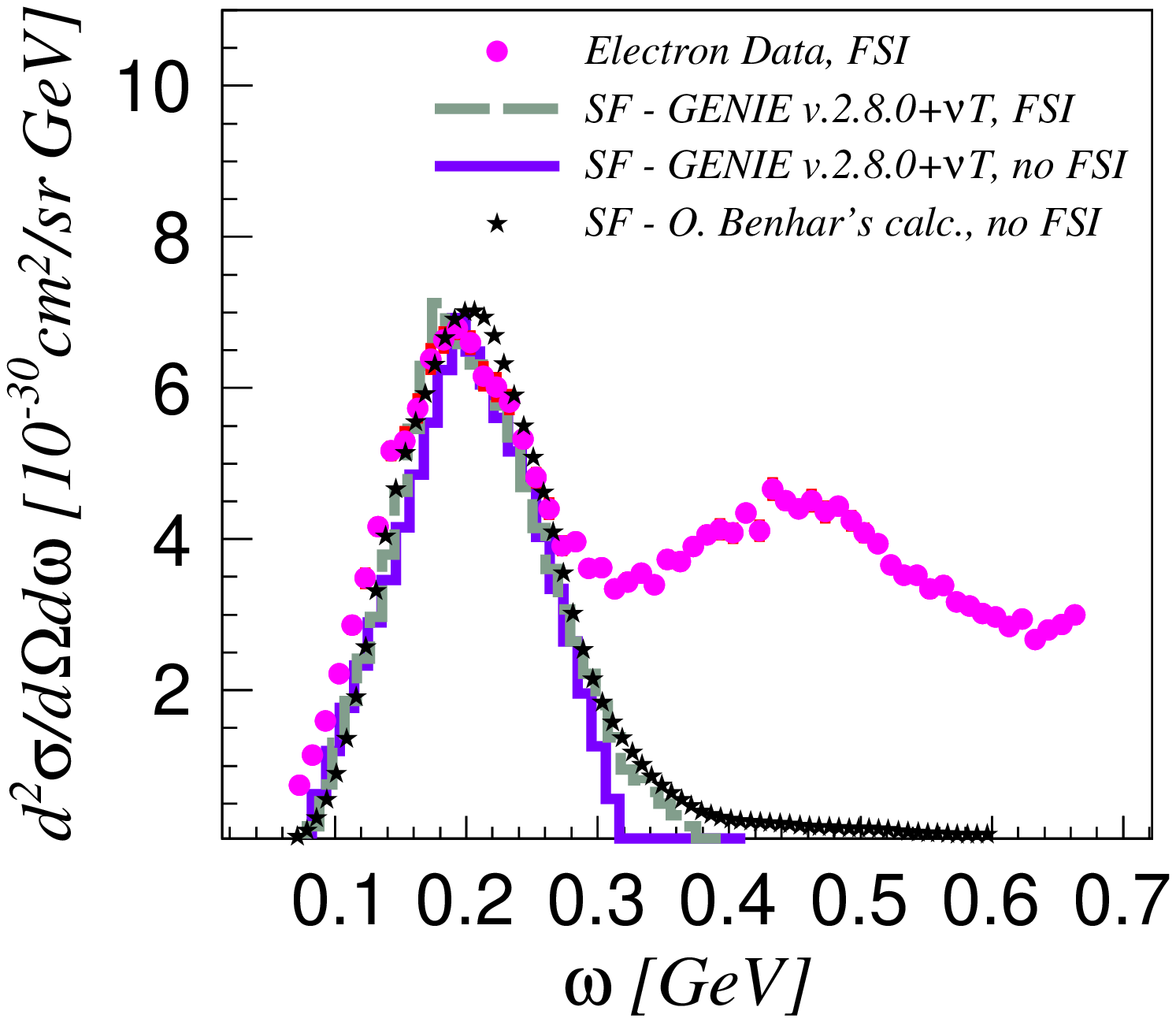}}
\subfigure[~$e+\textrm{C}\to e' + X$, $E_{e}=1.299$~GeV, $\theta_{e}=37.5$~deg]{\includegraphics[width=0.9\columnwidth]{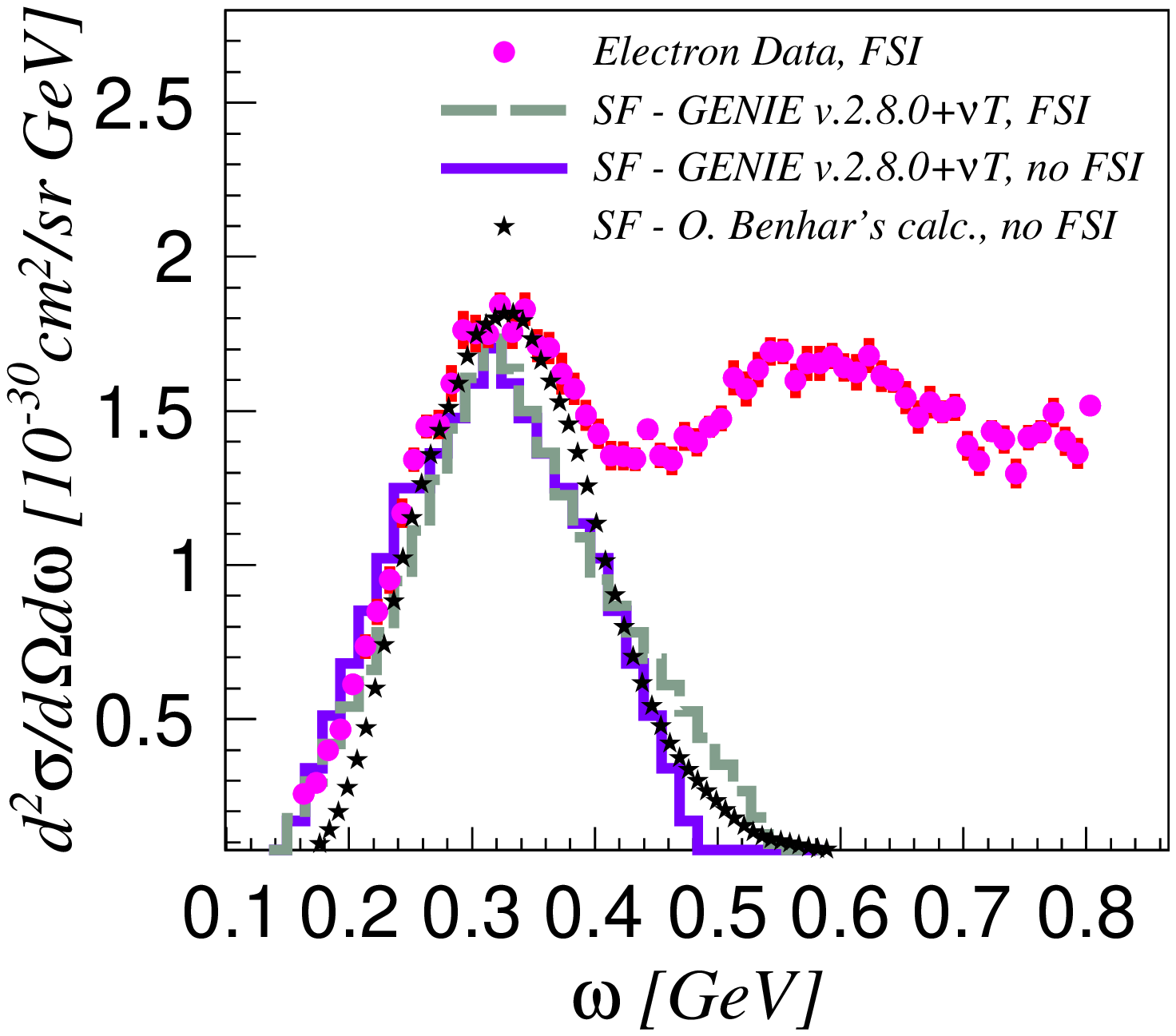}}
\subfigure[~$e+\textrm{Ca} \to e' + X$, $E_{e}=0.841$~GeV, $\theta_{e}=45.5$~deg]{\includegraphics[width=0.9\columnwidth]{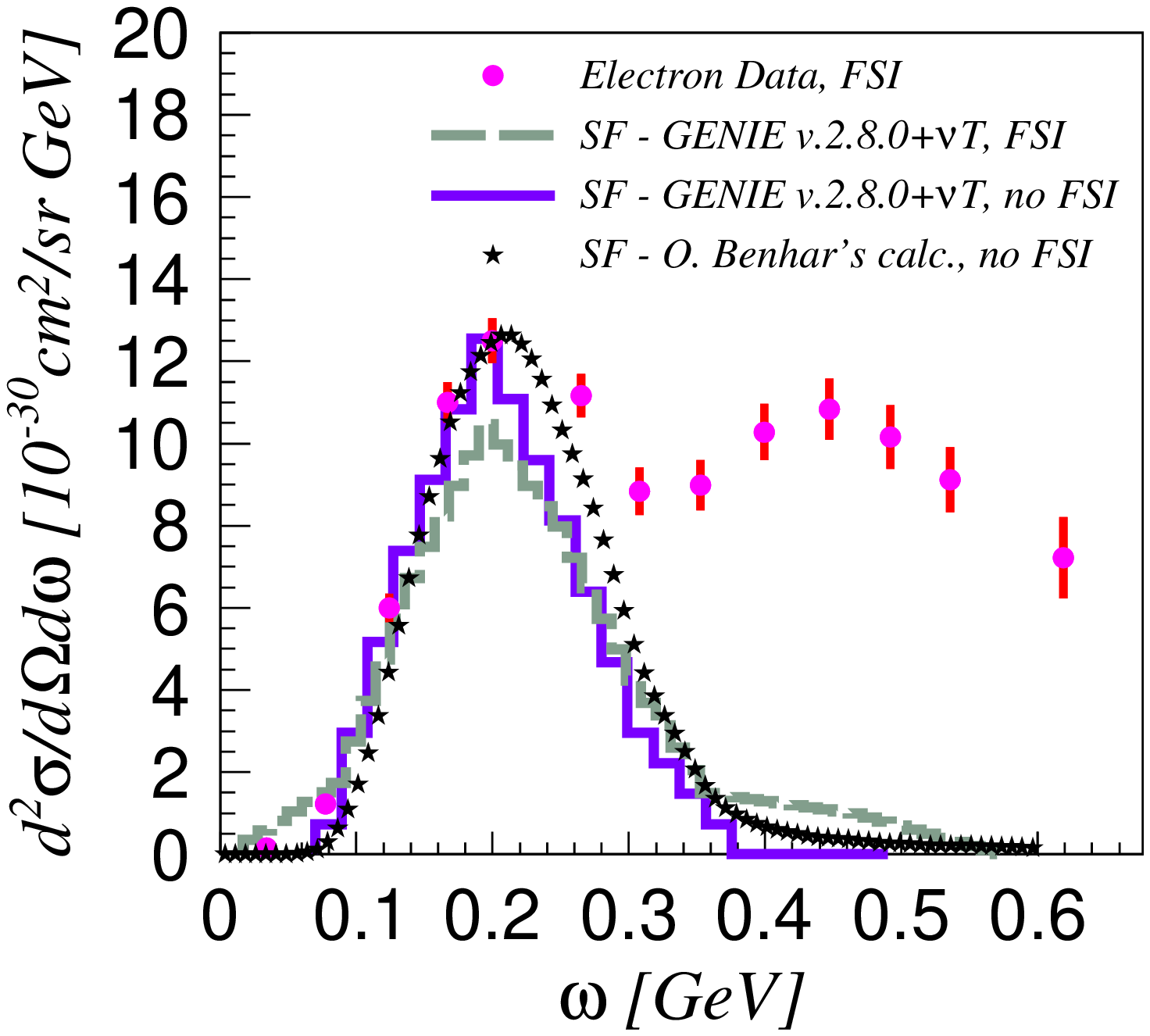}}
\subfigure[~$e+\textrm{Ar} \to e' + X$, $E_{e}=0.700$~GeV, $\theta_{e}=32$~deg]{\includegraphics[width=0.9\columnwidth]{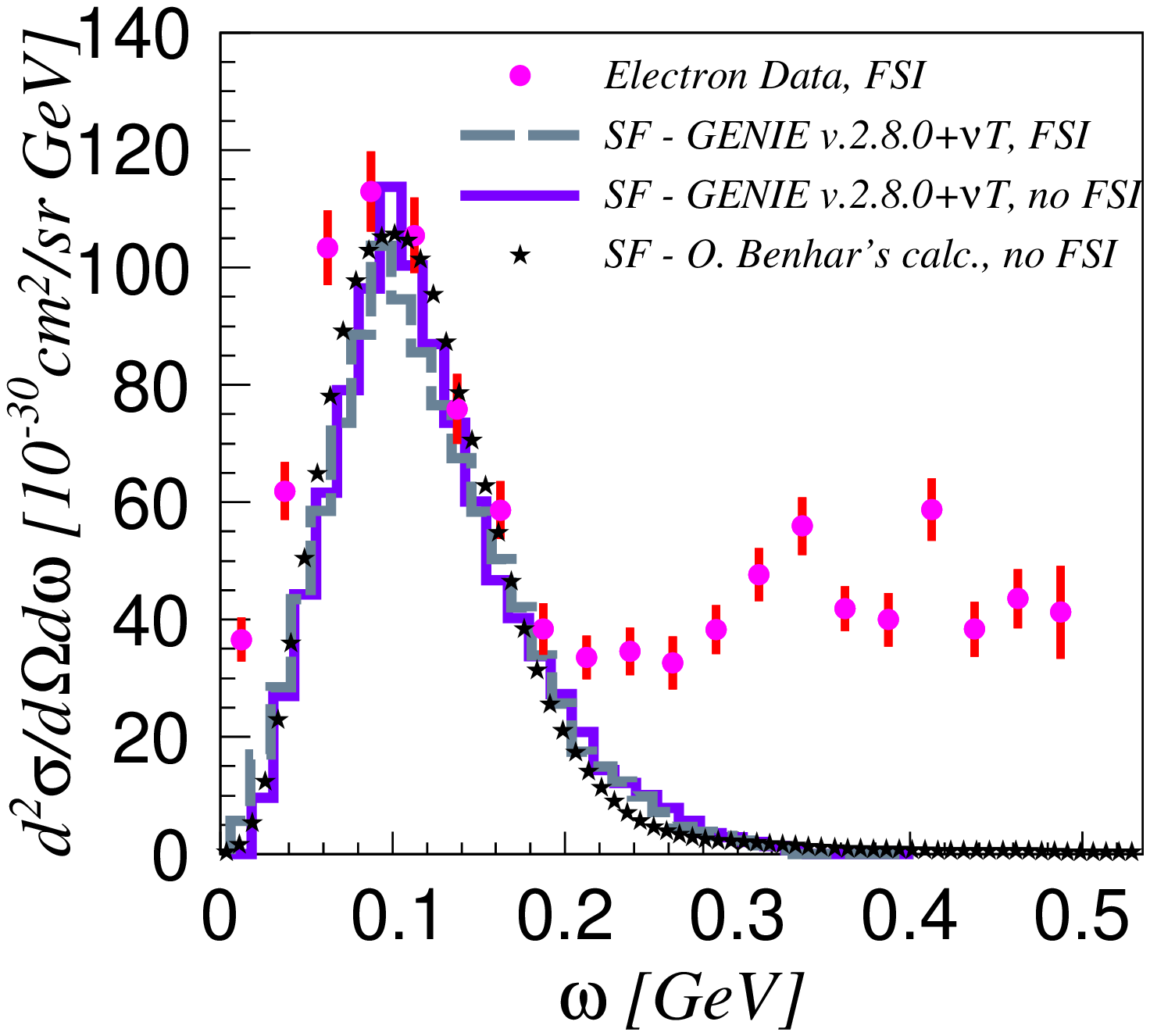}}
\caption{(Color online).~Double differential electron-nucleus cross section in the quasielastic channel. The curves labeled SF have been obtained using Eqs.~\eqref{MC1}--\eqref{MC3} and the model spectral functions of Refs.~\cite{LDA} (for carbon) and~\cite{Ankowski:2007uy} (for calcium and argon).
The data are taken from Refs.~\cite{12C2} (for carbon),~\cite{Williamson:1997} (for calcium), and~\cite{Anghinolfi:95n} (for argon). Carbon and calcium data  are available online at {\color{blue} {\protect \url{http://faculty.virginia.edu/qes-archive/index.html}}.}\label{F11}}
\end{center}
\end{figure*}

\vspace{0.3cm}

\section{The GENIE Event Generator}
\label{subgenie}

\vspace{0.2cm}
The GENIE event generator, in its latest official release, $2.8.0$, provides the
simulation of CCQE neutrino interactions within two different nuclear models: the RFGM and
the spectral function (SF) approach. In addition to the CCQE channel, both nuclear
models can be used to simulate interactions leading to different hadronic final states,
such as resonance production and decay, pion production and deep-inelastic scattering.
A detailed description of the treatment of these processes can be found in
Refs.~\cite{Andreopoulos:2009rq,Dytman:2011zza}.

The SF implementation in the official
GENIE $2.8.0$ does not include either calcium or argon, and the algorithm used to sample the
nucleon energy-momentum distribution is different from the one employed in our work.

To carry out the simulation following the scheme outlined in the previous Section,
we have developed a few modules to replace those of the GENIE $2.8.0$ package.
The developed modules, to which we will refer as the $\nu T$ package, are not part of any GENIE release (official or development).
However, they are compatible with the GENIE $2.8.0$ official release, and will be shortly available online at the Virginia Tech website. In what follows, we will use the GENIE $2.8.0$ official release with the additional $\nu T$ modules and refer to this code as GENIE $2.8.0+\nu T$.

The modifications introduced in the $\nu T$ package will be analyzed in the following Sections. In Fig.~\ref{F11} we show a comparison between the results of our simulations of $2 \times 10^6$ events and the measured electron scattering
cross sections, for different targets and kinematical setups, meant to validate our implementation.

\begin{figure}
\includegraphics[width=\columnwidth]{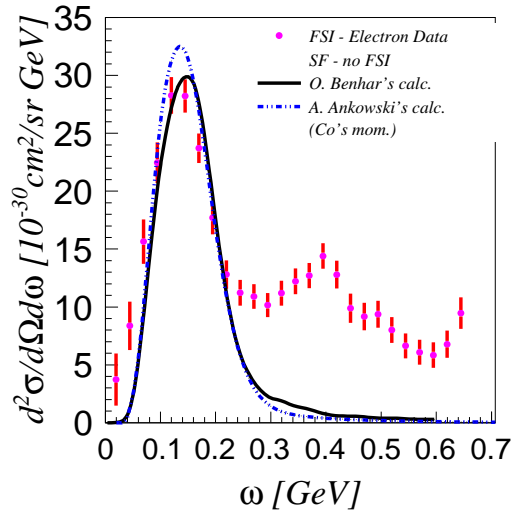}
\caption{(Color online). Inclusive electron-oxygen cross section at $E_e = 0.880 \ {\rm GeV}$ and $\theta_e = 32 \ {\rm deg}$, computed using the LDA \cite{LDA} (solid lines) and GSF \cite{Ankowski:2007uy} (dot-dash lines). \label{compare1}}
\end{figure}
\begin{figure}
\includegraphics[width=\columnwidth]{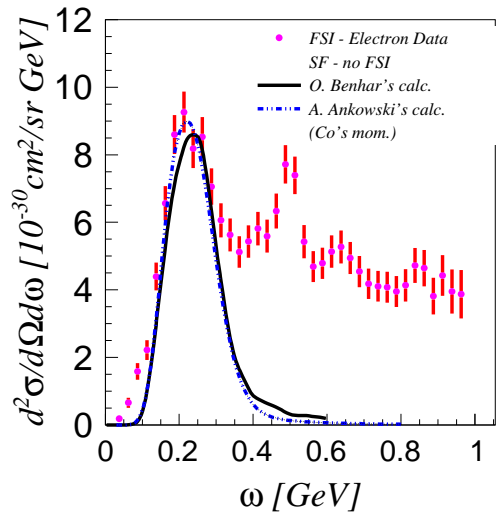}
\caption{(Color online). Same as in Fig.\ref{compare1}, but for $E_e~=~1.2 \ {\rm GeV}$ and $\theta_e = 32 \ {\rm deg}$. \label{compare2}}
\end{figure}

Carbon events have been generated using the spectral function of Ref.~\cite{LDA}, while for calcium and argon we employ the spectral functions of Ref.~\cite{Ankowski:2007uy}. Note that, in order to allow for a consistent comparison with the cross sections computed using Eqs.~\eqref{MC1}--\eqref{MC3},
for GENIE $2.8.0+\nu T$  we show results obtained with and without inclusion of FSI effects~\cite{Dytman:2014}.

The overall agreement of the results obtained neglecting FSI is quite good, the small differences being largely ascribable to numerical accuracy.

In inclusive processes, FSI are known to lead to a shift of the energy loss distribution,
arising from interactions between the struck nucleon and the mean field of the spectators, and a redistribution
of the strength from the peak of the quasi-free bump to its tails, arising from rescattering processes. These features, more pronounced in heavier nuclei, can
be observed in the the GENIE $2.8.0+\nu T$ results presented in Fig.~\ref{F11}.
To compare the LDA and GSF models of the target spectral function, in Figs. ~\ref{compare1} and \ref{compare2} we report, as an example,
the inclusive electron-oxygen cross section for two selected kinematics. It appears that, while the widths of the quasi free bump, determined by the
Fermi momentum, are quite similar, discrepancies are observed in both the position of the maximum and its height, dictated by the energy dependence
of the spectral function.
\subsection{Sampling of $(p,E)$ pairs and $Q^2$ selection}

As mentioned above, there are differences between our numerical implementation of the spectral function and the one used in GENIE $2.8.0$. The main new feature is the sampling of the nucleon momentum and energy, $p$ and $E$, from the {\em two-dimensional} probability distribution $4 \pi p^2 P(p,E)$. This procedure, the results of which are illustrated in Fig.~\ref{fig:F00a}, turns out to be very efficient, as it exploits the strong energy-momentum correlation exhibited by the spectral function.
In GENIE $2.8.0$, on the other hand, the values of $p$ and $E$ are obtained applying the acceptance-rejection method to randomly generated pairs. This procedures amounts to treat $P(p,E)$ as a function of two uncorrelated variables. Besides, GENIE $2.8.0$ does not include spectral function models of either calcium or argon.

In addition to the sampling of the spectral function, we have modified the determination of the squared four momentum transfer.


In GENIE $2.8.0$, the value of $Q^2$ is selected randomly within a range defined by a set of minimum and maximum values, which can be tuned manually, and it is approved if passes the acceptance-rejection test based on the pre-calculated differential cross section $d\sigma/dQ^2$. The $Q^2$ selection is unaffected by the initial-state nucleon kinematics, which is dictated by the dynamical model employed to describe the target ground state, RFGM or SF.

As it clearly appears in Eq.~\eqref{def:vertex}, however, a consistent implementation of the IA scheme requires that, while the  tensor $L_{\mu\nu}$ is determined from the lepton
kinematical variables $k$ and $q = k-k^\prime$ only, the nucleon
tensor depends on the initial nucleon momentum, ${\bf p}$ and ${\widetilde q} \equiv (\widetilde \omega,{\bf q})$, which in turn depends on the removal
energy $E$ through its time component, defined by Eq. \eqref{omega:tilde}.

The main original features of GENIE $2.8.0+\nu T$ can be summarized as follows:
%
\begin{itemize}
\item because part of the energy transfer to the target, $\delta \omega = \omega -\widetilde \omega$,  goes into excitation energy of the spectator particles
[see Eqs. \eqref{delta:omega} and \eqref{omega:tilde2}],  the energy transfer to the interacting nucleon is $\widetilde \omega < \omega$;

\item the value of $Q^2$ is constrained further requiring that the scattering process be QE, i.e. imposing the condition
\[
(p + q)^2 = m^2 \ ,
\]
where the initial four-momentum of the the interacting nucleon is $p \equiv(M_A - E_{A-1}, {\bf p})$, with  $E_{A-1}~=~\sqrt{ (M_A - m + E)^2 + {\bf p}^2 }$.
This requirement obviously implies that the initial-state nucleon kinematics affects the $Q^2$ selection process.
\end{itemize}

\begin{figure}[htbn!]
\includegraphics[width=1.2\columnwidth]{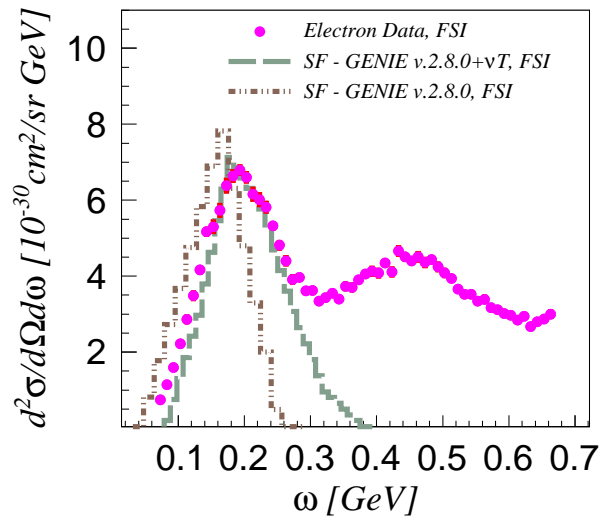}
\caption{(Color online).~Comparison between the measured inclusive electron-carbon cross section at beam energy E$_e$=0.961~GeV and scattering angle $\theta_e$=37.5~deg
\cite{12C1} and the result obtained from GENIE $2.8.0$ (dot-dash line)
and GENIE $2.8.0+\nu T$ (dashes).}
\label{fig:electron_data_comparison}
\end{figure}

\begin{figure}[htbn!]
\includegraphics[width=1.2\columnwidth]{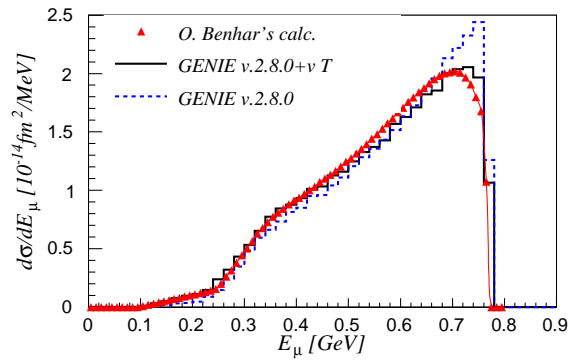}
\caption{(Color online).~Outgoing muon momentum distribution corresponding to muon neutrinos of 0.8~GeV scattering off an oxygen target. The results obtained using  GENIE $2.8.0$ and GENIE $2.8.0+\nu T$ are represented by the dotted and solid line, respectively. The theoretical results from Ref.~\cite{PhysRevD.72.053005} (labeled O. Benhar's calc.)  are shown by triangles. The effect of Pauli blocking is included in all three calculations, while FSI are not taken into account.}
\label{fig:muon_neutrino_comparison}
\end{figure}

In the end, the selected $Q^2$ satisfies the relation
\begin{equation}
 Q^2 = |{\bf q}|^{2} - (E_{\nu}-E_{\ell})^{2} = |{\bf q}|^{2} - \omega^{2} \ ,
\end{equation}
where $E_{\ell}$ is the outgoing lepton's energy, while $|{\bf q}|$ and $\omega$ are the magnitude of the three-momentum transfer and
the energy transfer, respectively. The additional constraint
\begin{equation}
\label{eq:q_boundary}
|{\bf k}| - |{\bf k}^\prime| \leq |\bf q| \leq |{\bf k}| + |{\bf k}^\prime| \ ,
\end{equation}%
where $|{\bf k}|=E_\nu$ and ${\bf k}^\prime = |{\bf k} + {\bf q}|$ is the three-momentum of the outgoing lepton,
is also applied to the generated $Q^2$ in order to enforce momentum conservation.
The effect of all the applied modification---sampling of the spectral function and determination of $Q^2$---is illustrated in Figs.~\ref{fig:electron_data_comparison} and~\ref{fig:muon_neutrino_comparison}.
In Fig.~\ref{fig:electron_data_comparison} we compare the results obtained using GENIE $2.8.0$ and $2.8.0+\nu T$ to electron scattering data for carbon at E$_e$=0.961~GeV and $\theta_e$=37.5~deg. It is apparent that GENIE $2.8.0$ fails to predict both position and width of the measured cross section, that are
dictated by the energy and momentum dependence of the spectral function, respectively.
Figure~\ref{fig:muon_neutrino_comparison} shows the  $E_\mu$-distribution corresponding to 800~MeV
muon neutrinos interacting with an oxygen target. The effect of Pauli blocking is taken into account following the procedure discussed in Ref.~\cite{PhysRevD.72.053005} and the results of Benhar {\em et al} \cite{PhysRevD.72.053005} are also displayed, for comparison.
%
%
\begin{figure*}[htbn!]
\subfigure[~GENIE $2.8.0$ with RFGM]{\includegraphics[width=0.68\columnwidth]{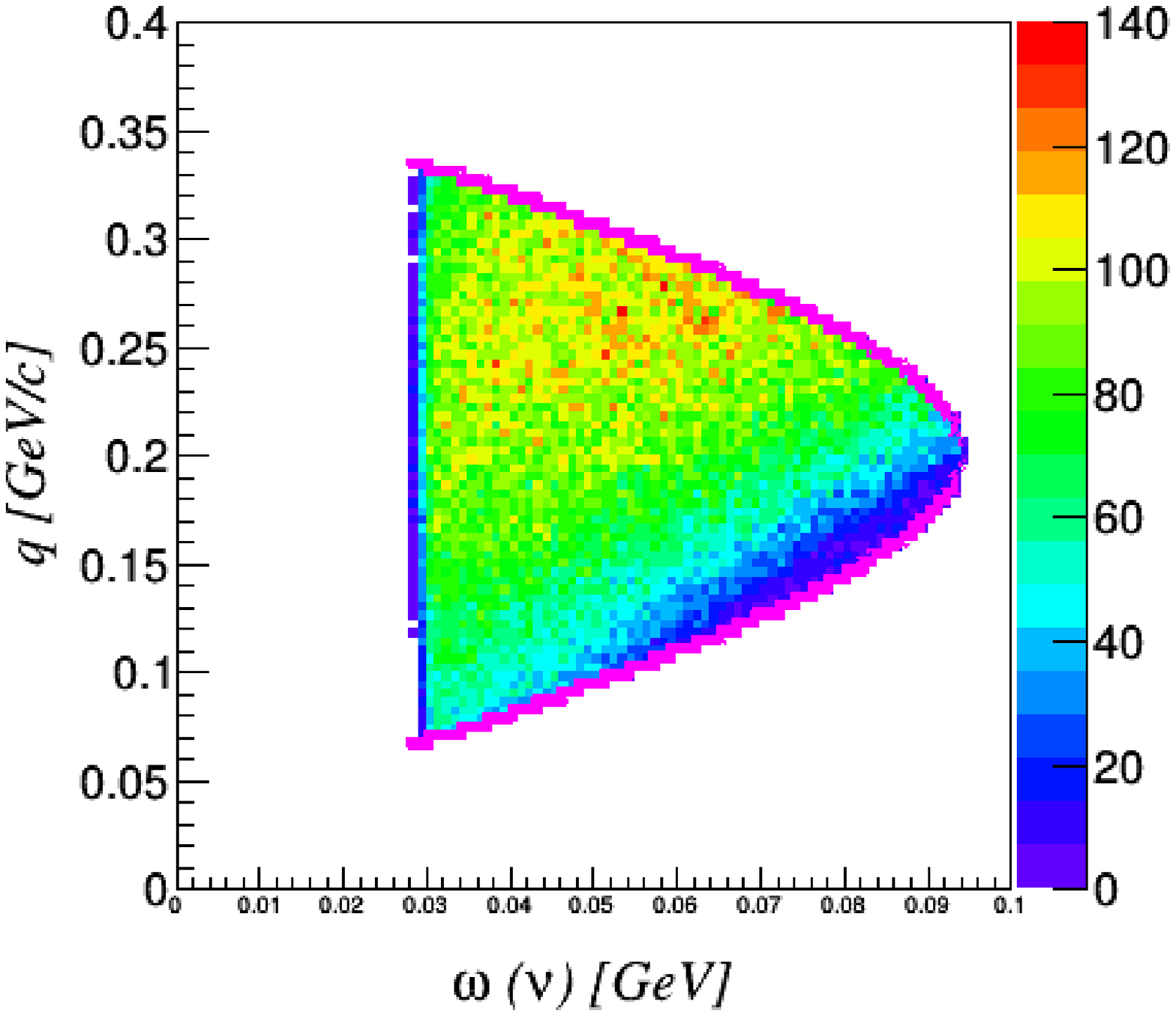}\label{fig:pse200a}}
\subfigure[~GENIE $2.8.0+\nu T$ with RFGM]{\includegraphics[width=0.68\columnwidth]{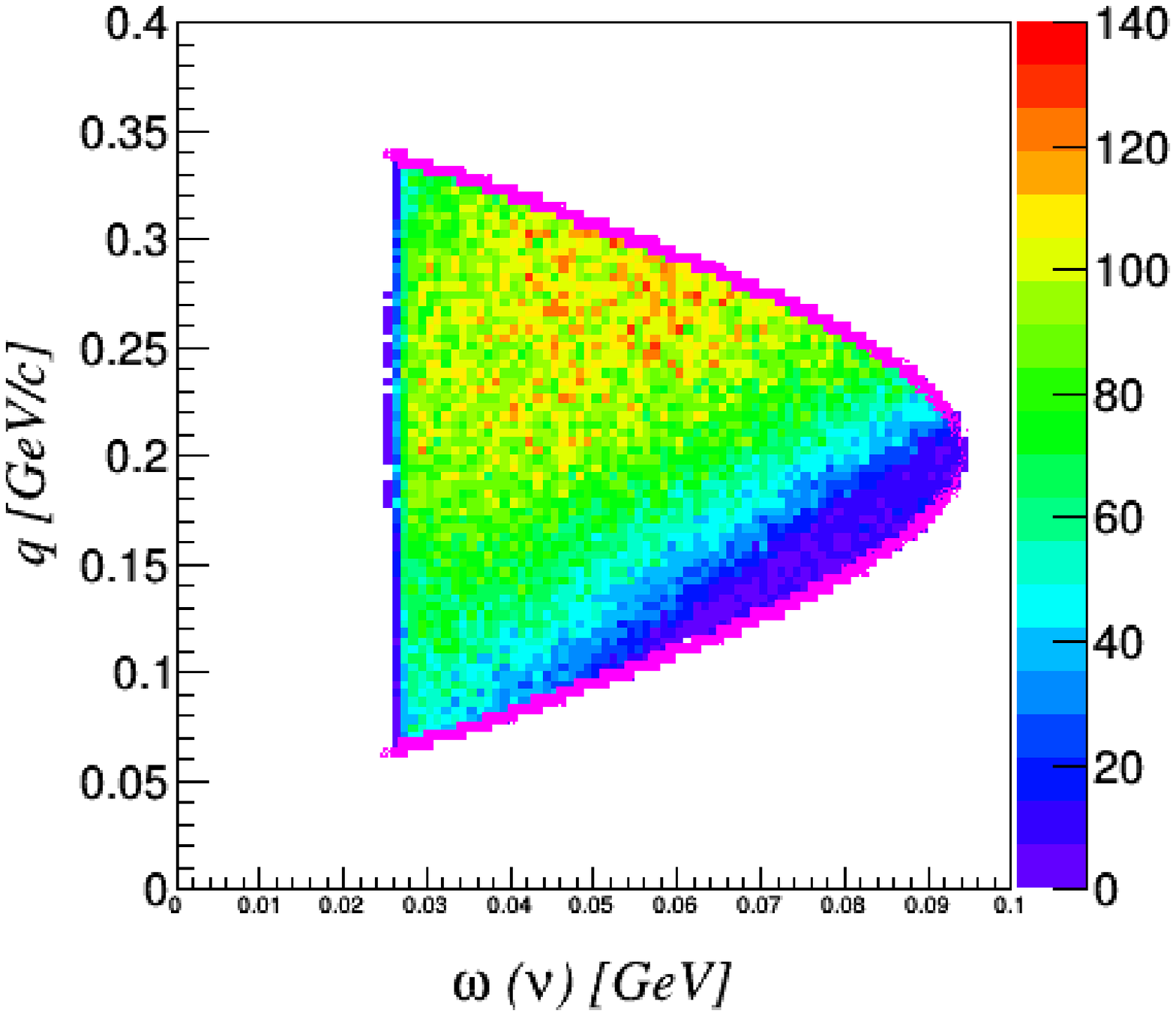}\label{fig:phase_space_e20_b}}		
\subfigure[~GENIE $2.8.0+\nu T$ with SF]{\includegraphics[width=0.68\columnwidth]{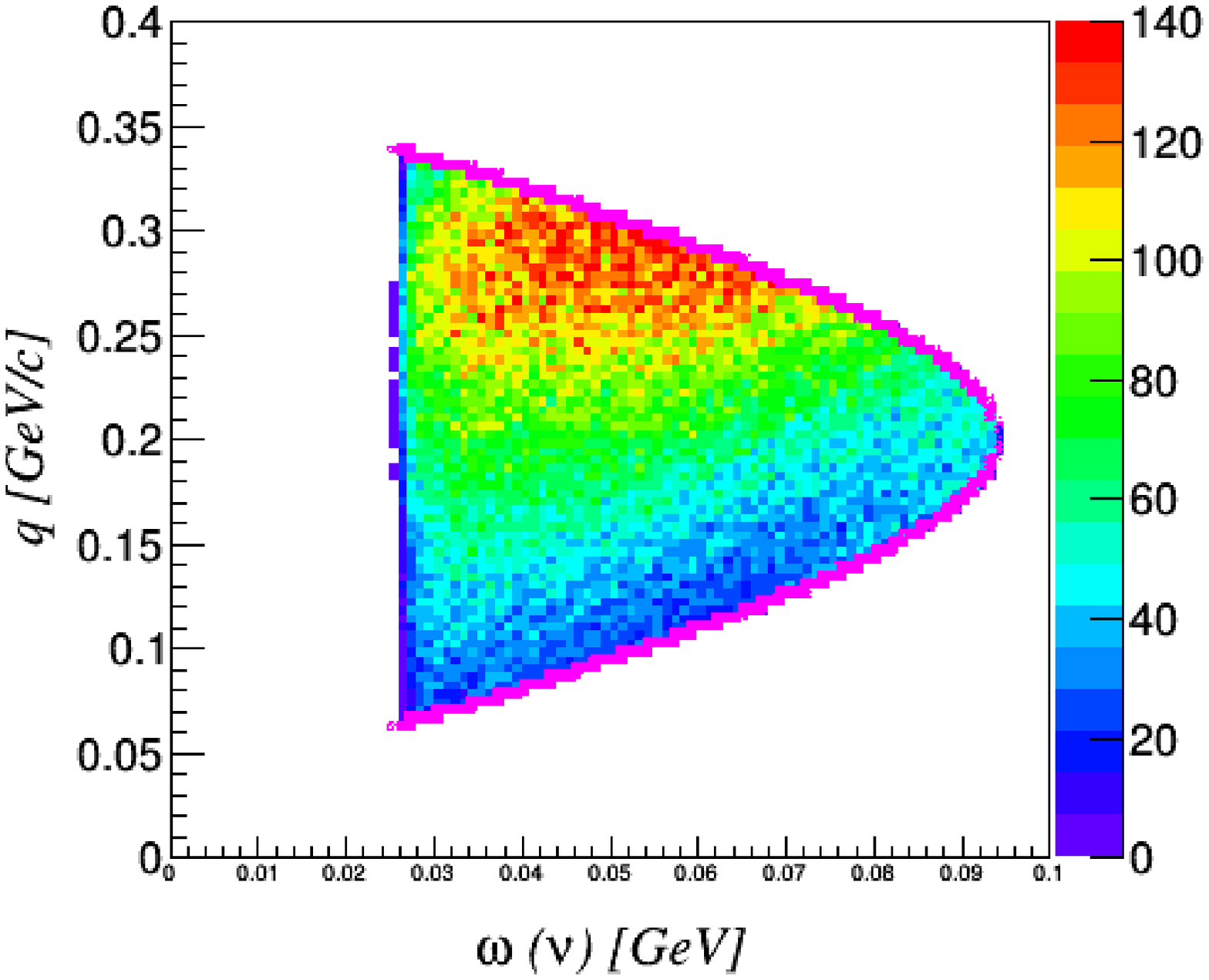}\label{fig:phase_space_e20_c}}
\caption{(Color online). Event distribution within kinematically allowed regions of the ($|{\bf q}|,\,\omega$) plane obtained at neutrino energy 200 MeV using GENIE $2.8.0$  with RFGM and GENIE $2.8.0+\nu T$ with both RFGM and SF.\label{fig:phase_space}}
\end{figure*}
\begin{figure}[htbn!]
\includegraphics[width=1.2\columnwidth]{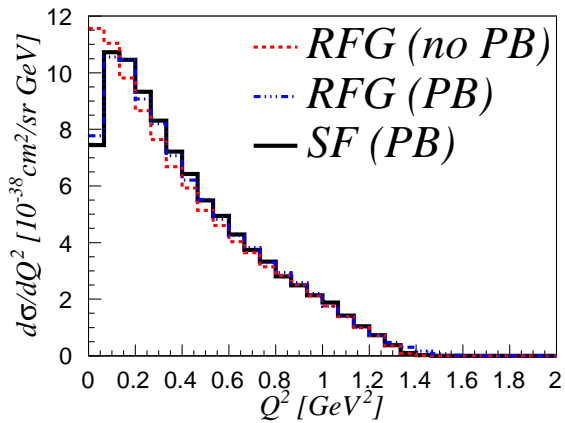}
\caption{(Color online).~Comparison of the differential cross sections $d\sigma / dQ^2$ for oxygen obtained from $2 \times 10^5$ CCQE events using RFGM, with
(dot-dash line) and without (dotted line) inclusion of Pauli blocking, respectively. and the SF model with Pauli blocking (solid line) at $E_\nu$ = 1~GeV.
For both the nuclear models (RFGM and SF) FSI are turned off. \label{fig:q2_xsec_oxygen}}
\end{figure}
\begin{figure*}[htbn!]
\subfigure[~$\nu+\textrm{O} \to \mu + X$, with no Pauli blocking, and no~FSI]{\includegraphics[width=0.95\columnwidth]{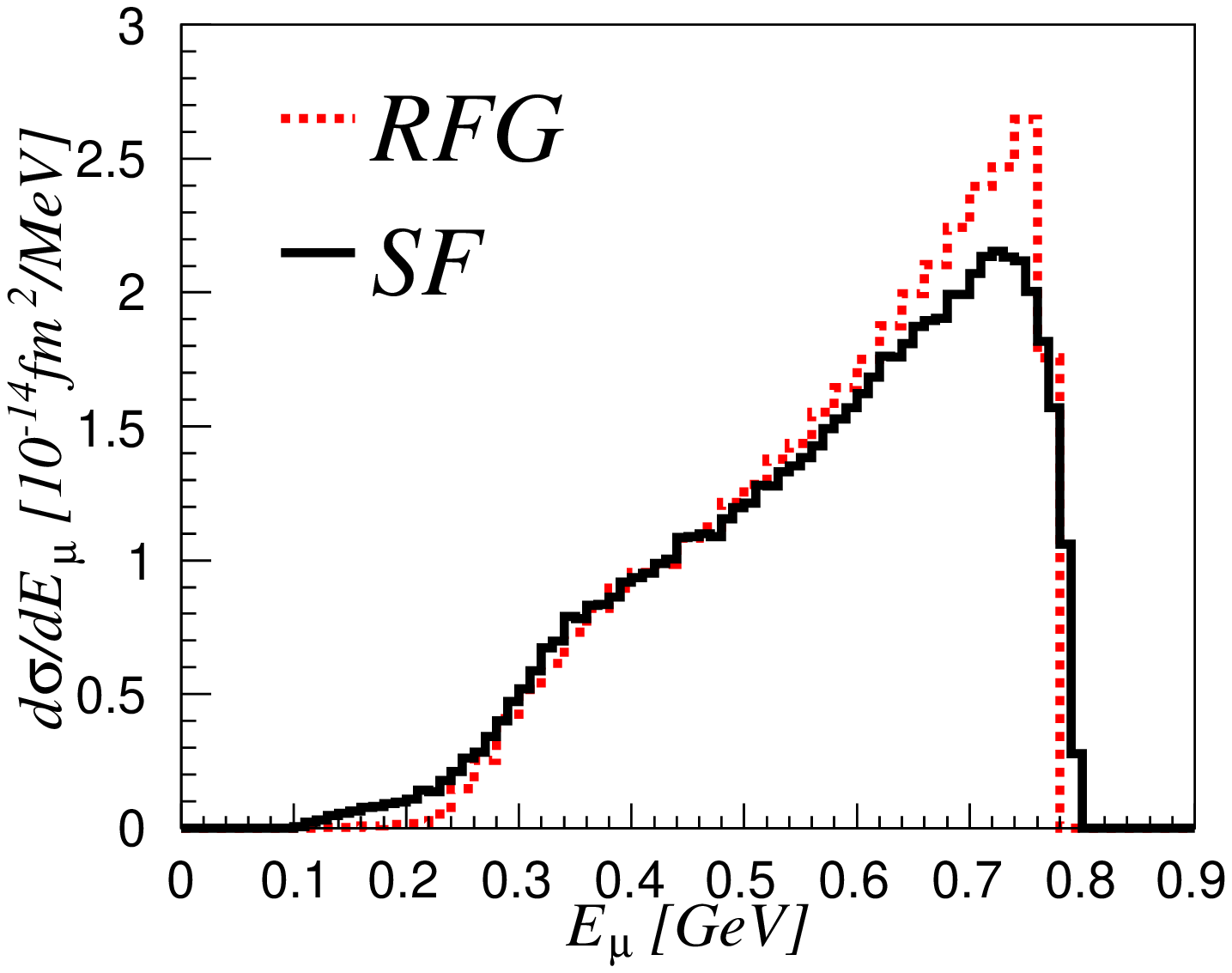}\label{fig:emu_xsec_a}}
\subfigure[~$\nu+\textrm{Ar} \to \mu + X$, with both Pauli blocking and FSI included]{\includegraphics[width=0.95\columnwidth]{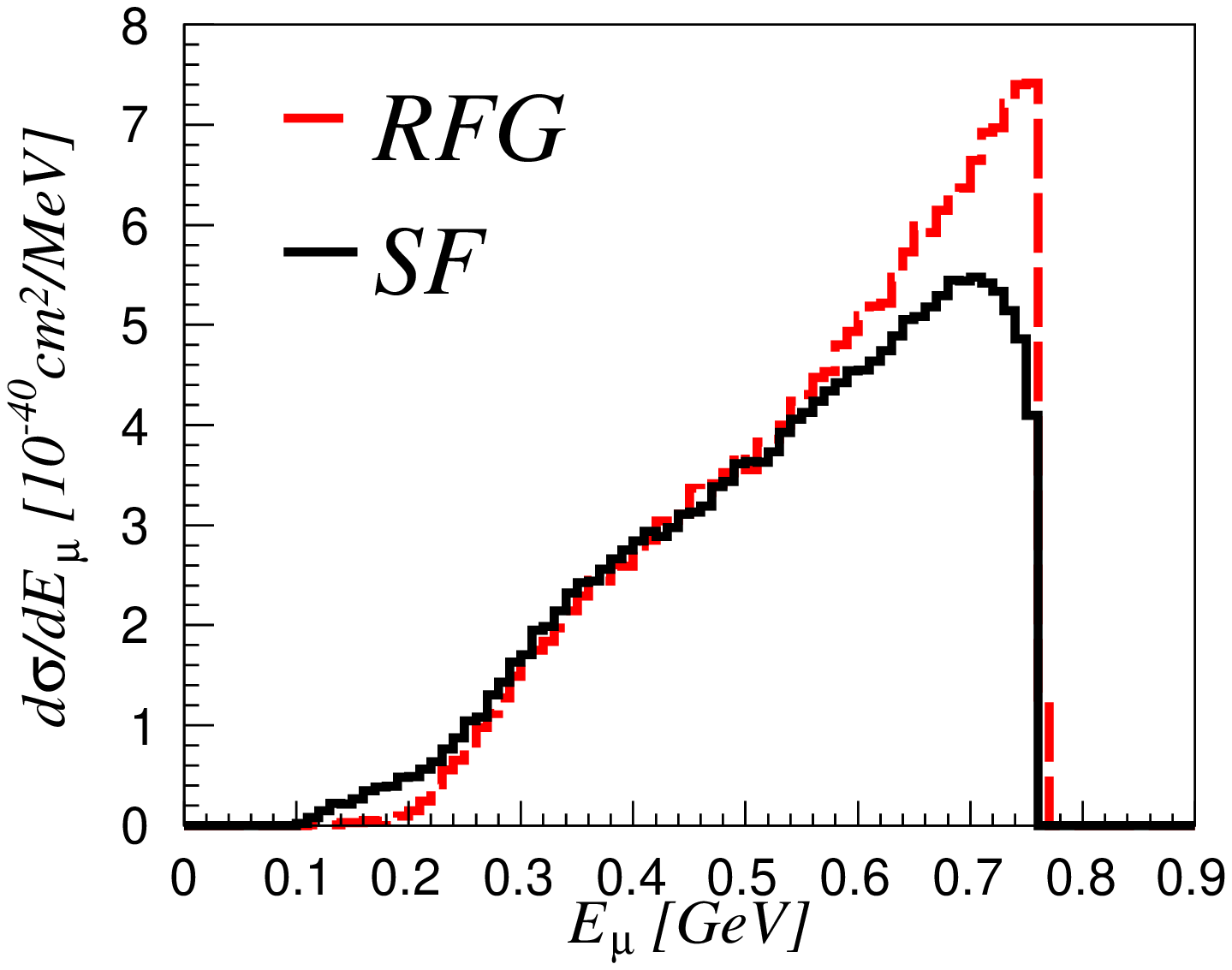}\label{fig:emu_xsec_b}}
\caption{(Color online).~ Comparison of the differential CCQE cross sections $d\sigma/dE_{\mu}$ of (a) oxygen and (b) argon at neutrino energy $E_\nu = 800$~MeV, obtained using GENIE $2.8.0+\nu T$ with RFGM and SF.\label{fig:emu_xsec}}
\end{figure*}

%
\par Figure~\ref{fig:phase_space} illustrates the kinematically allowed phase space for muon neutrinos of energy 200 MeV, obtained within RFGM and SF with the boundaries determined by Eq.~\eqref{eq:q_boundary}, as well as the corresponding event distribution. To show the effects of the modified $Q^2$ selection, in Figs.~\ref{fig:pse200a} and~\ref{fig:phase_space_e20_b},
we compare the RFGM results obtained using GENIE $2.8.0$ and $2.8.0+\nu T$, respectively. It is apparent that GENIE $2.8.0+\nu T$ is less likely to select the points close to the boundaries at low energy transfer and low momentum transfer.

On the other hand, in the SF model, nucleon-nucleon correlations lead to the occurrence of larger number of events at higher values of the momentum and energy transfers. We illustrate it in Fig.~\ref{fig:phase_space_e20_c}, making use of the implementation of the SF approach in GENIE $2.8.0+\nu T$.
\subsection{Lepton Kinematics}
As pointed out above, the value of the energy transfer at the interaction vertex depends on both momentum, $p$, and removal energy, $E$, of the struck nucleon, the distribution of which, dictated by the nucleon spectral function, is illustrated in Fig.~\ref{fig:F00a}.
As a consequence, the $(p,\,E)$ distribution also affects the kinematical variables of the outgoing lepton, i.e. its energy and scattering angle relative to the direction of the incoming neutrino.
In this Section, we show the different shapes of the distributions of the lepton kinematical variables obtained from RFGM and SF, reflecting the different underlying models of nuclear dynamics.
Figure~\ref{fig:q2_xsec_oxygen} shows a comparison of the $Q^2$ distributions of $2 \times 10^5$ CCQE events with $E_\nu = 1$~GeV in oxygen, obtained using RFGM and SF.
The Fermi momentum and average separation energy employed in the RFGM calculation are $p_{F}=209$~MeV and $\epsilon_0 = 27$~MeV, respectively. Note that the SF result has been obtained taking into account Pauli blocking of the momentum of the final state nucleon, leading to the rejection of most events corresponding to $Q^2<0.2$~GeV$^{2}$, following the procedure of Ref.~\cite{PhysRevD.72.053005}. The RFGM distributions have been obtained both with and without inclusion of Pauli blocking.
The cross section as a function of energy of the outgoing muon is displayed in Fig.~\ref{fig:emu_xsec}. The neutrino energy is $E_\nu = 800$ MeV, and  panels (a) and (b) correspond to oxygen and argon, respectively.
A discrepancy between RFGM and SF in the number of scattered leptons at the highest lepton energy, corresponding to the lowest energy transfer, is clearly visible.
Note that, in addition to the quenching at large $E_\mu$, the SF distribution exhibits a tail extending to very low muon energy.
These events, corresponding to large $\omega$, become kinematically allowed in the presence of nucleon-nucleon correlations, as illustrated in Fig.~\ref{fig:phase_space_e20_c}. Pauli blocking and FSI, included in the argon results, have been neglected in oxygen. Comparison between the results of panels (a) and (b) shows that the main features of the distributions are not strongly affected by these effects.
\subsection{Reconstructed kinematics}
\begin{figure*}[htbn!]
\subfigure[~$E_{\nu}=0.2$~GeV]{\includegraphics[width=0.9\columnwidth]{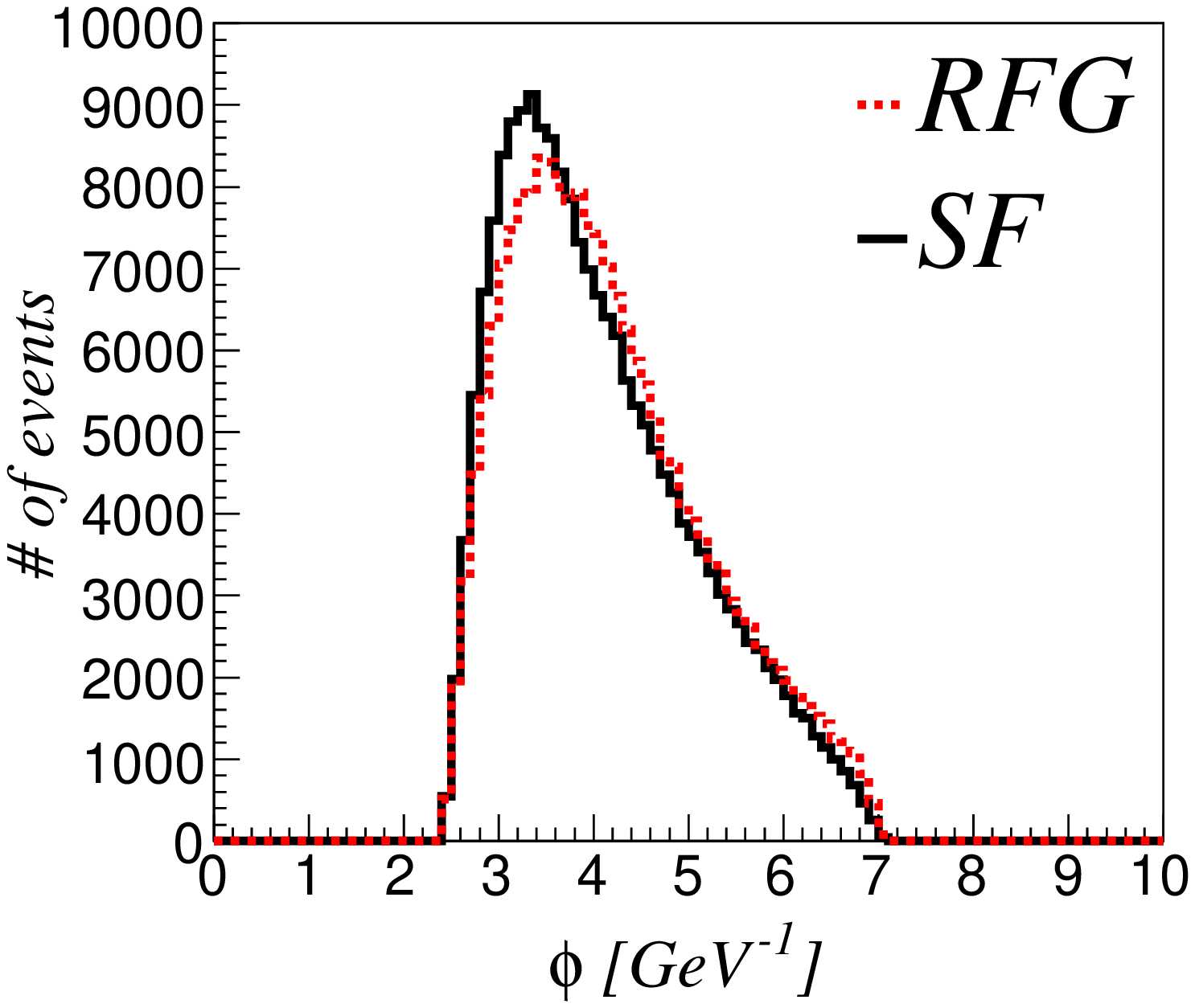}}
\subfigure[~$E_{\nu}=0.3$~GeV]{\includegraphics[width=0.9\columnwidth]{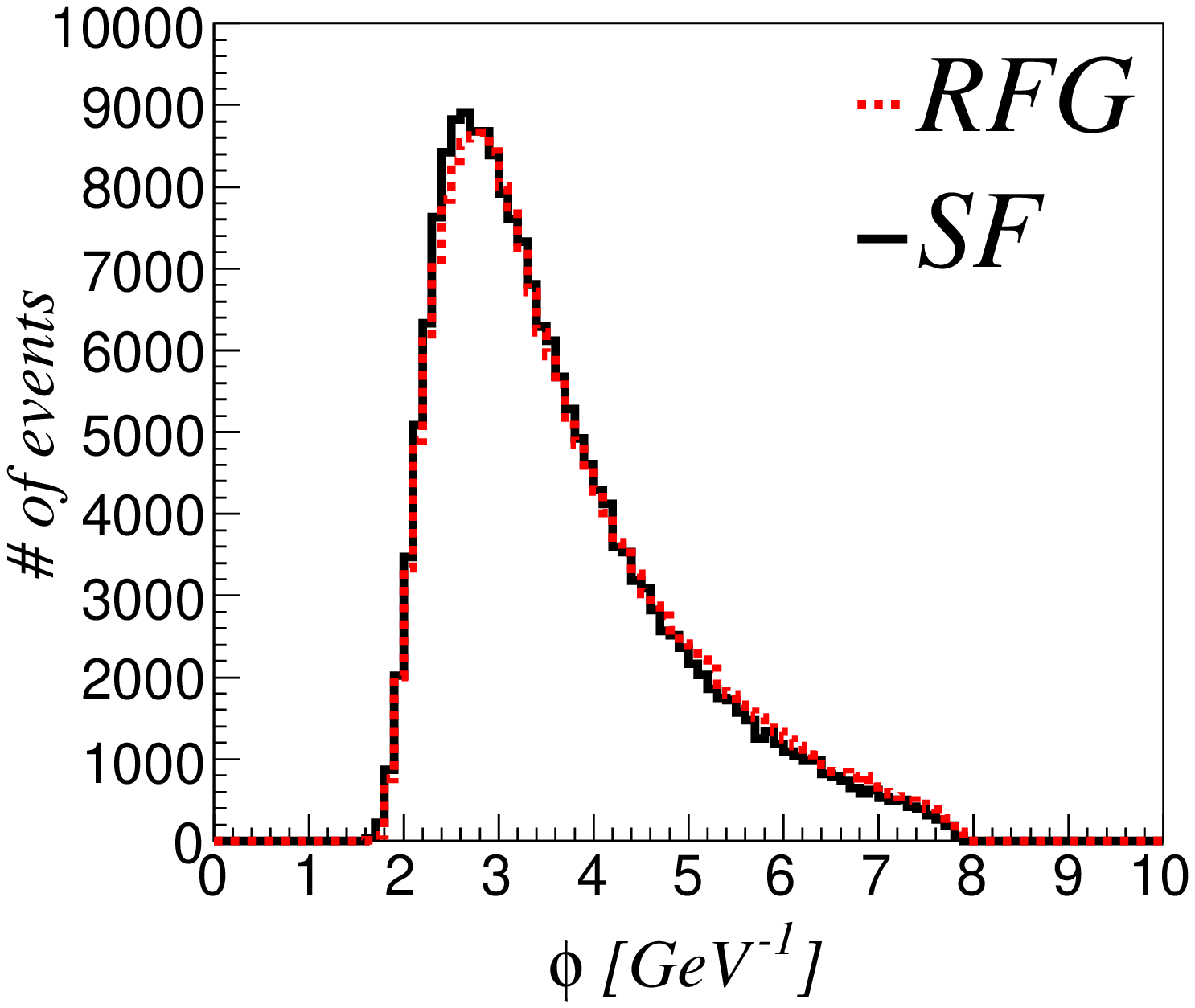}}
\caption{(Color online).~$\phi$ distribution for scattered muon neutrinos of energy (a) $E_{\nu}=0.2$~GeV and (b) 0.3~GeV on an oxygen target, obtained using GENIE $2.8.0+\nu T$ with RFGM and SF without considering the effect of Pauli Blocking and neglecting FSI. \label{fig:phi_0.2_0.6}}
\end{figure*}

In this Section, we will discuss the reconstruction of neutrino energy and $Q^2$ in CCQE processes. We will use two variables, $\beta$ and $\phi$, first proposed in Ref.~\cite{Ankowski:2010yh}. They are defined in terms of the observed kinematical variables,  $|{\bf k}^\prime|$ and the scattering angle of the outgoing lepton relative to the beam direction in the lab frame, $\theta_{\mu}$, as
\begin{equation}
\label{eq:beta}
 \beta = E_{\mu} - |{\bf k}'| \cos \theta_{\mu} \ ,
\end{equation}
where $E_\mu = \sqrt{ |{\bf k}'| + m_\mu^2 }$, $m_\mu$ being the muon mass, and
\begin{equation}
\label{eq:phi}
 \phi  = \frac{1}{m_{\mu}+\beta} \ .
\end{equation}
The $\phi$ distributions of events generated in oxygen and corresponding to neutrino energy 200~MeV and 300~MeV, displayed in Fig.~\ref{fig:phi_0.2_0.6} (a) and (b),
respectively, show that $\phi$ is generally in the range $1~\leq~\phi~\leq~10$/GeV.
The reconstructed neutrino energy and $Q^2$ can be expressed in terms of $\beta$ according to
\begin{equation}
\label{eq:reconstructed_enu}
 E^{\rm rec}_{\nu} = \frac{E_{\mu}(M_{n}-\epsilon_0)-(\epsilon_0^{2}-2M_{n}\epsilon_0+m^{2}_{\mu}+\Delta M^{2})}{2(M_{n}-\epsilon_0-\beta)} ,
\end{equation}
and
\begin{equation}
\label{eq:reconstructed_q2}
 Q^2_{\rm rec} = -m^{2}_{\mu} + 2E^{\rm rec}_{\nu}\beta,
\end{equation}
where $\Delta M^{2} = M^{2}_{n}-M^{2}_{p}$, $M_{n}$ and $M_{p}$ being the neutron and proton masses, respectively, while ${\epsilon_0}$  denotes the
average binding energy of the struck neutron.
\par The variable $\beta$ can be used to identify unphysical CCQE event with a negative value of the reconstructed energy.
From Eq.~\eqref{eq:reconstructed_enu}, it follows that such events correspond to $\beta \gtrsim 0.9$~GeV, implying in turn $\phi \lesssim 1$/GeV. Note that the amount of unphysical reconstructed events is reduced by $\sim$50\% with the improved determination of the lepton kinematics implemented into GENIE $2.8.0+\nu T$.
%
%
\par
The $\phi$ distributions of Fig.~\ref{fig:phi_0.2_0.6} also illustrate the difference between RFGM and SF, that turn
out to become negligible for the larger neutrino energy.
Owing to the nontrivial bias associated with the reconstruction process~\cite{Ankowski:2010yh},
the reconstructed kinematic quantities are not the best choice as independent variables
for the differential cross section.
\par
\begin{figure}[htbn!]
\includegraphics[width=0.9\columnwidth]{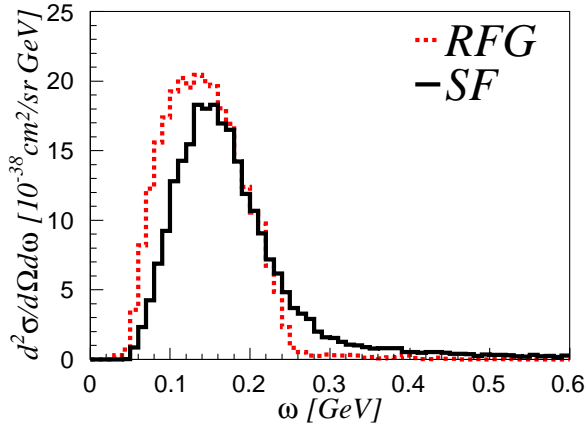}
\caption{(Color online).~Double differential CCQE cross sections of muon neutrino of 1~GeV and scattering angle 30~deg on an oxygen target obtained using RFGM and SF. Pauli blocking effect is in this case included, but the effect of FSI is still neglected.\label{fig:muon_kine_xsec_oxygen}}
\end{figure}

 Measured kinematical variables, such as the muon kinetic energy, $T_\mu$,  and scattering angle,  $\theta_\mu$, provide
a much more reliable option.
As an example, Fig.~\ref{fig:muon_kine_xsec_oxygen} shows
the oxygen CCQE double differential cross section at beam energy 1~GeV, plotted as
a function of the energy loss. It clearly appears that nucleon-nucleon correlations, included
in the SF calculation, move strength from the region of the quasielastic bump to higher
values of the energy loss $\omega$. Obviously, this mechanism leads to the appearance of
muons of low kinetic energy, as shown in Fig.~\ref{fig:muon_kine_xsec_oxygen}.
%
%
%

\begin{figure*}[htbn!]
\subfigure[~RFGM]{\includegraphics[width=\columnwidth]{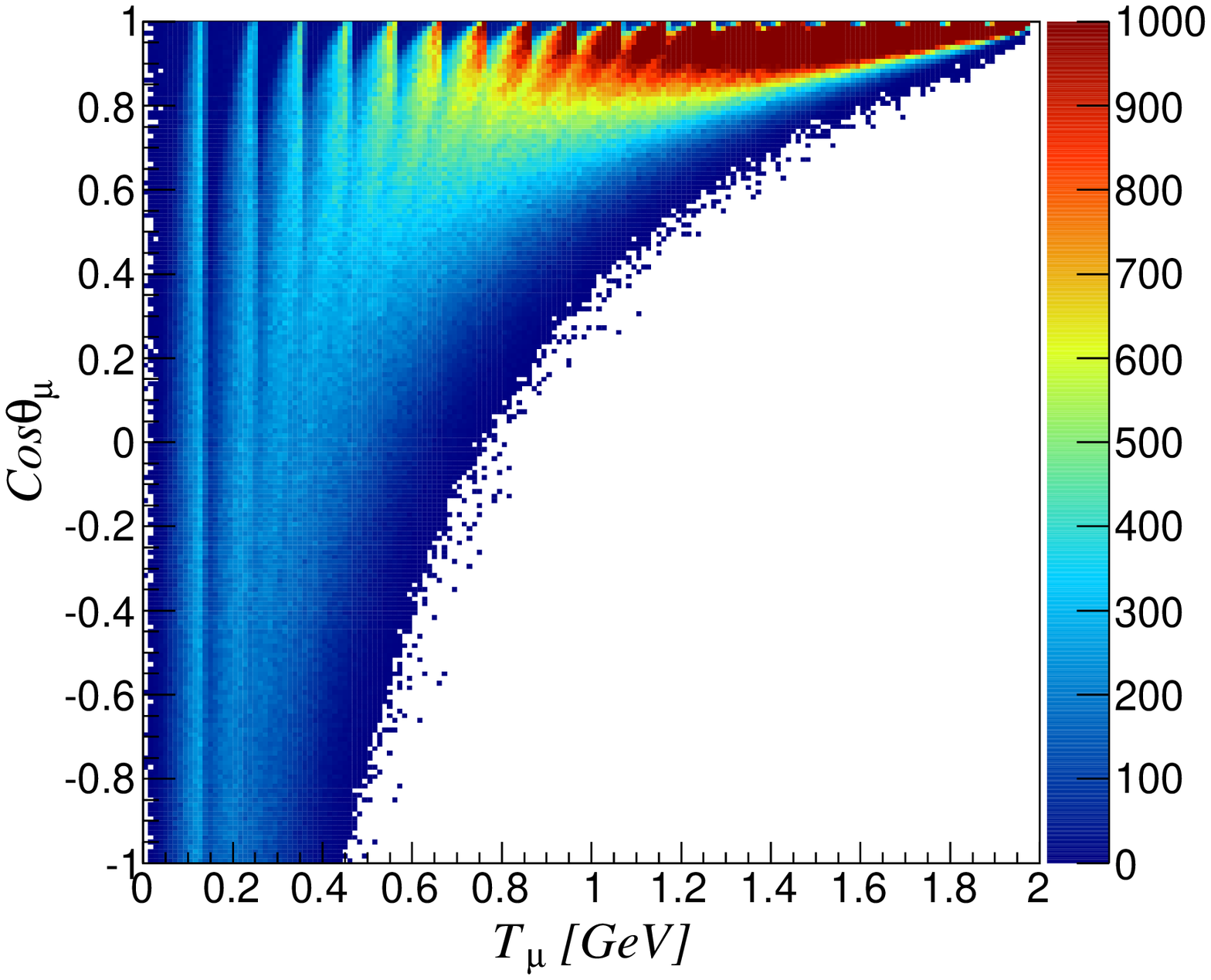}}
\subfigure[~SF]{\includegraphics[width=\columnwidth]{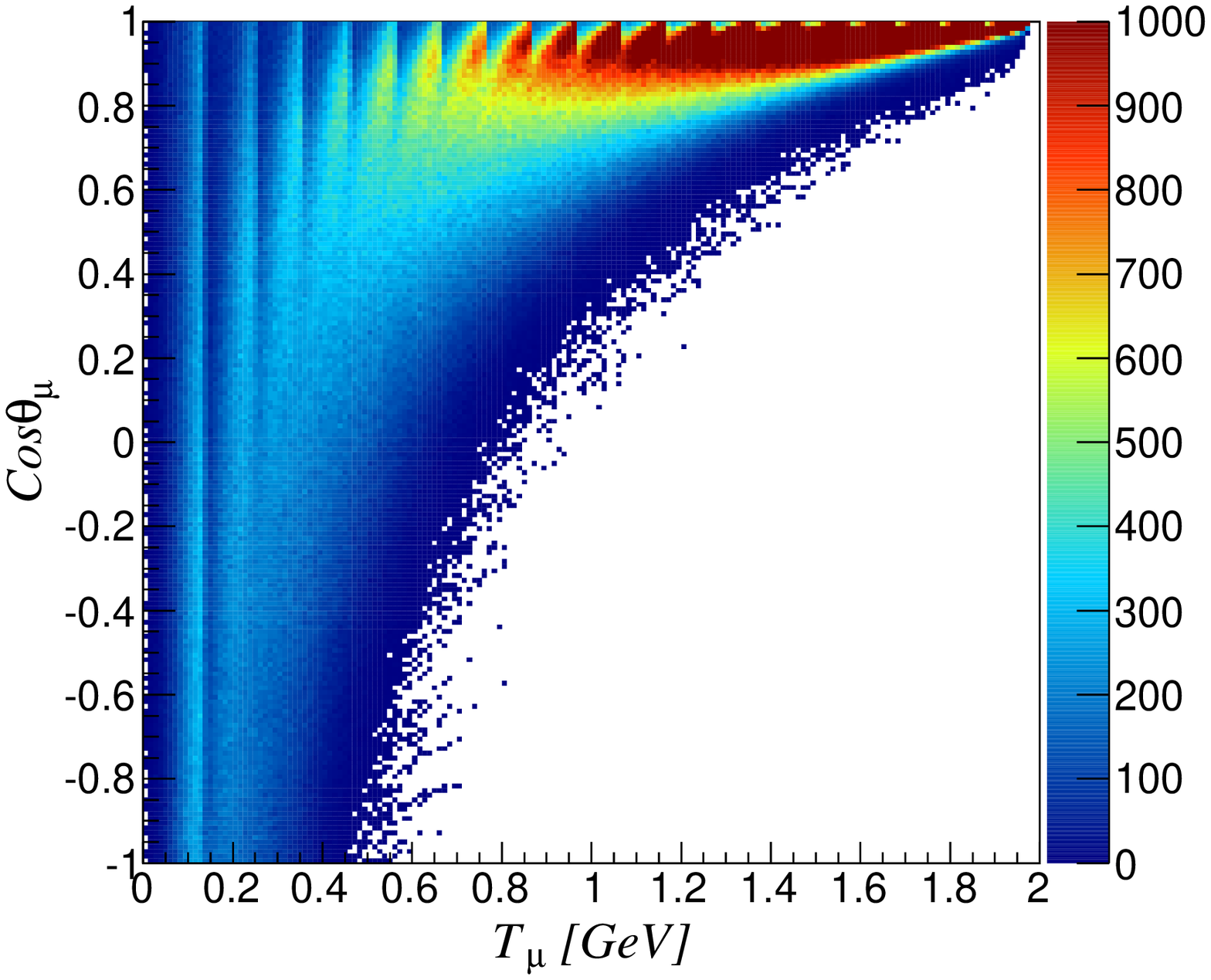}}
\caption{(Color online). CCQE event distribution for $\nu_\mu$ interactions with oxygen generated using GENIE $2.8.0+\nu T$ with (a) RFGM  and (b) SF for a uniform neutrino flux of energy between 0.2 and 2.0 GeV.  \label{fig:mu_kine_oxygen}}
\end{figure*}

As the probe energy is not known in neutrino scattering, the measurement of $T_\mu$ does not provide the information
on the energy transfer $\omega$. Figure~\ref{fig:mu_kine_oxygen} shows on $(T_\mu,\cos \theta_\mu)$ plane the distribution of CCQE events for muon neutrino scattering off the oxygen target. The calculations for a uniform neutrino energy distribution, ranging from 200~MeV to 2~GeV in bins of 100~MeV, have been carried out using GENIE $2.8.0+\nu T$ with both RFGM and SF, generating $2\times10^5$ events at each value of energy.
%
%
\section{Effect on neutrino oscillations\label{sec:oscillation}}
In this Section, we describe an analysis aimed at gauging the influence of the description of neutrino interactions on the extraction of oscillation parameters.
For this purpose, we consider a typical  $\nu_\mu$ disappearance experiment, consisting of two identical detectors of fiducial volume 1.0~kton and 22.5~kton, placed 1.0~km and  295.0~km from the neutrino beam production point, respectively.
Both detectors use carbon ($^{12}$C) as nuclear target and they have identical properties in terms of energy resolution and detector efficiencies.
The experiment is assumed to take data for 5 years with a 750~kW beam power, and the flux is set to be that of the T2K experiment. The setup, summarized in Table~\ref{tab:exp_setup},
is the same as the one used in Refs.~\cite{Huber:2009cw,Coloma:2013rqa,Coloma:2013tba}. We have used GENIE $2.8.0$ to simulate all but CCQE interactions,  for which we have used alternatively GENIE $2.8.0$ and our modified version of GENIE $2.8.0+\nu T$.
\begin{table*}[ht!]
\centering
\caption{Experimental setup used for the oscillation analysis presented in this work~\cite{Coloma:2013tba}.}
\vspace{0.5cm}
\begin{tabular}{l@{\quad}c@{\quad}c@{\quad}c@{\quad}c@{\quad}c@{\quad}}
\hline \hline
 & Baseline  & Fid. mass & Flux peak & Beam Power & Run. time \\ \hline
Far     &  295~km    & 22.5~kt  & \multirow{2}{*}{0.6~GeV}  &  \multirow{2}{*}{750~kW}  & \multirow{2}{*}{5~yrs}  \cr
Near       &  1.0~km     & 1.0~kt    & 			&		  &			 \\ \hline\hline
\end{tabular}
\label{tab:exp_setup}
\end{table*}
It should be noted that this setup is largely simplified with respect to a real experiment so our conclusions should be regarded as a \emph{lower} limit on what the impact of different nuclear interaction models and  numerical implementations would be in a real experiment.
Following Refs.~\cite{Huber:2009cw,Coloma:2013tba} we perform the oscillation analysis using the GLoBES sensitivity framework~\cite{globes1,globes2}.
\par The oscillation parameters used in our analysis are
\begin{align}
\nonumber
&\Delta m^2_{21}  =  7.64  \times 10^{-5}  \ \textrm{eV}^2 \ \ , \ \ \Delta m^2_{31}\  = 2.45\times 10^{-3} \ \textrm{eV}^2 \ ,\nonumber \\
&\theta_{12}  =  33.2 \ {\rm deg} \ , \  \theta_{23}  =  45 \ {\rm deg} \ , \ \theta_{13}  = 9  \   \textrm{deg} \  , \ \delta = 0 \ ,
\label{eq:oscparams}
\end{align}
and we only focus on the determination of the so-called atmospheric parameters: $\theta_{23}$ and $\Delta m^2_{31}$.
\par We consider the effect of the following different nuclear models on the determination of the atmospheric neutrino oscillation parameters:
\begin{itemize}
\item RFGM with the original Q$^2$ selection as in GENIE $2.8.0$
\item RFGM with the new Q$^2$ selection discussed in Sec.~\ref{subgenie}, as in GENIE $2.8.0+\nu T$
\item SF for $^{12}$C, as in GENIE $2.8.0+\nu T$.
\end{itemize}
\par
\begin{figure}[htbn!]
\includegraphics[width=\columnwidth]{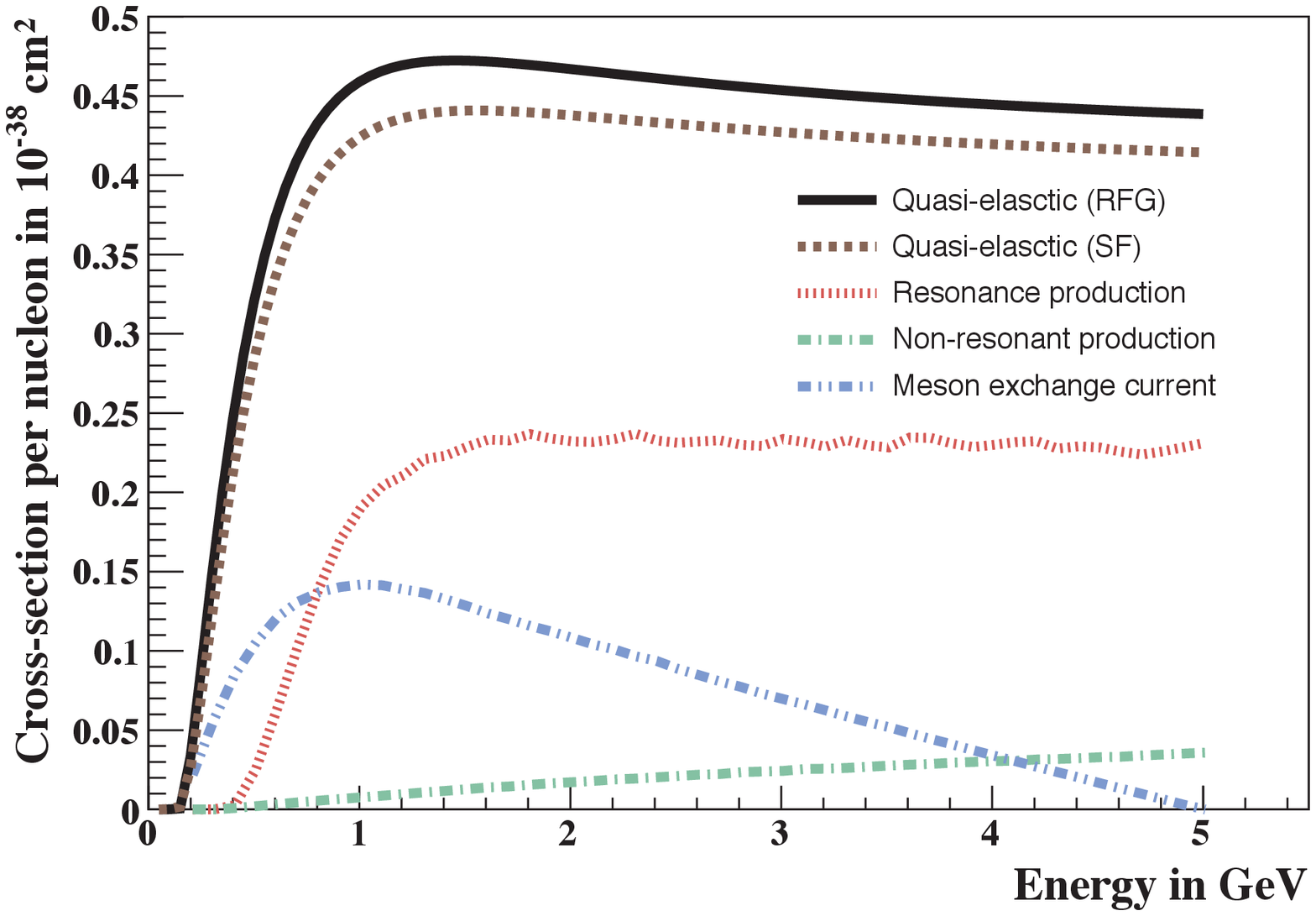}
\caption{(Color online).~QE and QE-like cross-sections per nucleon in $^{12}$C as a function of neutrino energy. Different curves represent different channels or different nuclear models used to simulate a particular channel.\label{fig:xsec}}
\end{figure}

 In the oscillation analysis we only consider events that are QE like, that is they contain no pions in the final state. In addition to the pure neutrino QE interactions, other channels included in the QE-like classification are resonant pion production (RES), non-resonant pion production (non-RES), and excitation of two-particle--two-hole final states through interactions involving meson-exchange currents (MEC/2p2h). The QE-like events due to the missing pion in the final state are mostly classified as QE and they are indeed indistinguishable from the pure QE events.
We generate RES, non-RES, and MEC/2p2h neutrino interactions using GENIE $2.8.0$,  while in the case of QE we use both GENIE $2.8.0$ and GENIE $2.8.0+\nu T$. A more detailed description of these interaction mechanisms can be found in Ref.~\cite{Coloma:2013tba}. We consider only neutrinos in the energy range of $0 < E_{\nu} < 2$~GeV. The contribution of deep inelastic scattering (DIS) and pion productions from high resonances (high-RES) at these energies is not very large, and becomes negligible once we require that the neutrino events have no pions in the final state.
The cross sections per nucleon on $^{12}$C for all QE-like cases listed above are shown in Fig.~\ref{fig:xsec} as a function of neutrino energy. It clearly appears that the effect of nuclear models on the QE cross section is large, the difference between SF and RFGM being $\sim$20\% . Similar results have been reported in Refs.~\cite{PhysRevD.72.053005,Benhar:2006nr}. The other interaction models\footnote{The 2p2h model currently implemented in GENIE $2.8.0$ is not consistent with the SF (1p1h) model described in Sec.~\ref{sec:theory} and implemented in GENIE $2.8.0+\nu T$.} are the ones present in the current GENIE release $2.8.0$ and they are described in details in Ref.~\cite{Andreopoulos:2009rq,Dytman:2011zza}.
%
%
\par
\begin{figure}[htbn!]
\includegraphics[width=\columnwidth]{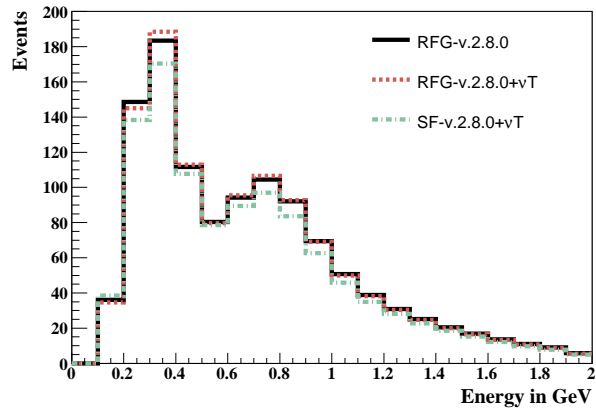}
\caption{(Color online).~Total event distributions as a function of the reconstructed neutrino energy for different nuclear models and Q$^2$ selection. The oscillation parameters have been set to their values in
Eq.~\eqref{eq:oscparams}, and  detection efficiencies have also been included. The neutrino energy is reconstructed assuming a pure QE events and according to Eq.~\eqref{eq:reconstructed_enu}.
\label{fig:events} }
\end{figure}

 The events numbers for all the QE-like mechanisms included in our oscillation analysis are summarized in Table~\ref{tab:qelike}.
%
\begin{table}[htbn!]
\centering
\caption{Number of events for the QE-like mechanisms included in the oscillation analysis performed in this work.}
\begin{tabular}{ | c | c | c | c | }
\hline
\hspace*{0.2cm} RES \hspace*{0.2cm}& \hspace*{0.2cm} non-RES \hspace*{0.2cm} & \hspace*{0.2cm} MEC/2p2h \hspace*{0.2cm} & \hspace*{0.2cm} Total QE-like \hspace*{0.2cm} \\
\hline
 173 & 8 &  231 & 412 \\
\hline
\end{tabular}
\label{tab:qelike}
\end{table}
%
The event numbers and event distributions as function of energy were generated using GLoBES. The background from neutral-current events was also generated using GLoBES, and found to consist of $\sim$254 events.
For the QE-like channels we also produced the migration matrices relating the true and reconstructed neutrino energies that were calculated using GENIE $2.8.0$ for all interactions but QE. The remaining pure QE rates were also computed with GLoBES, using inputs produced by GENIE $2.8.0$ and GENIE $2.8.0+\nu T$ with RFGM or SF. The number of events per interaction mode are summarized in Table~\ref{tab:qe}, while the signal distributions are shown in Fig.~\ref{fig:events}.
\par $M_{ij}\equiv N(E^{rec}_i, E^{true}_j)$ is defined as a migration matrix and it represents the probability that an event with a true neutrino energy in the bin $j$ ends up being reconstructed in the energy bin $i$. We reconstruct the neutrino energy for all QE-like events assuming a pure QE neutrino interaction as in Eq.~\eqref{eq:reconstructed_enu}. The migration matrices used in this work were produced using both GENIE $2.8.0$ and GENIE $2.8.0+\nu T$. All the migration matrices produced and used in our oscillation analysis are shown in Appendix~\ref{app:matrices}. Each matrix was produced considering 200,000 interactions for each of the true neutrino energy bins. We use bins of 100~MeV between 0 and 2~GeV and we considered only events with no-pion in final state.
The signal events are further corrected for the energy dependent detection efficiencies after the events are migrated to reconstructed neutrino energies , as described in more details in Ref.~\cite{Huber:2009cw}.
\par The QE only event distributions and the resonance, non-resonance and MEC/2p2h event distributions as function of reconstructed neutrino energy are shown respectively in Fig.~\ref{fig:ev2} and in Fig.~\ref{fig:ev1} in the Appendix~\ref{app:matrices}. In both Figs.~\ref{fig:ev2} and~\ref{fig:ev1} the oscillation parameters have been set to their values as in Eq.~\eqref{eq:oscparams}, and they are corrected for the detection efficiencies as well.
\begin{table}[htbn!]
\centering
\caption{Number of events for the QE-like neutrino interaction modes for the different nuclear models considered in our oscillation analysis.}
\begin{tabular}{ | c | c | c | c | }
\hline
\hspace*{0.2cm}  \hspace*{0.2cm}& \hspace*{0.2cm} Pure QE \hspace*{0.2cm} & \hspace*{0.2cm} QE-like \hspace*{0.2cm}  & \hspace*{0.2cm} Total \hspace*{0.2cm}  \\
\hline
 \hspace*{0.2cm}  RFGM $2.8.0$  \hspace*{0.2cm}    &  730  &  412 &  1142 \\
 RFGM $2.8.0+\nu T$  &  731  &  412 &  1143 \\
 SF $2.8.0+\nu T$     &  654  &  412 &  1066 \\
\hline
\end{tabular}
\label{tab:qe}
\end{table}
%
%
To evaluate the impact of three different simulation conditions (RFGM, RFGM + new $Q^2$ selection and SF) we took the event rates computed using GLoBES, applied to them the migration matrices computed with one particular setting of the neutrino interaction generator, and try to fit them using the matrices obtained with a different setting. By doing this, the possible biases on the oscillation parameters, induced  by the different nuclear models or $Q^2$ selection introduced in GENIE $2.8.0+\nu T$, can be quantified in a robust fashion. The effects of the RFG and SF models of GENIE $2.8.0+\nu T$ were also compared. We recall that the main focus of our analysis is the extraction of the atmospheric parameters through the disappearance of $\nu_\mu$. The $\chi^2$ utilized exploits both the rate and spectral distortion of the event distributions. Its functional form is the same as in Ref.~\cite{Coloma:2013tba}.
\par The atmospheric parameters used as an input in our oscillation analysis are
\begin{eqnarray}
\theta_{23}  =  45 \ {\rm deg} \  ,\  \ \ \Delta m^2_{31}  = 2.45\times 10^{-3} \ \textrm{eV}^2 \nonumber  \,
\end{eqnarray}
and the remaining parameters were held fixed during the fit.
\par
\begin{figure}[htbn!]
\includegraphics[width=\columnwidth]{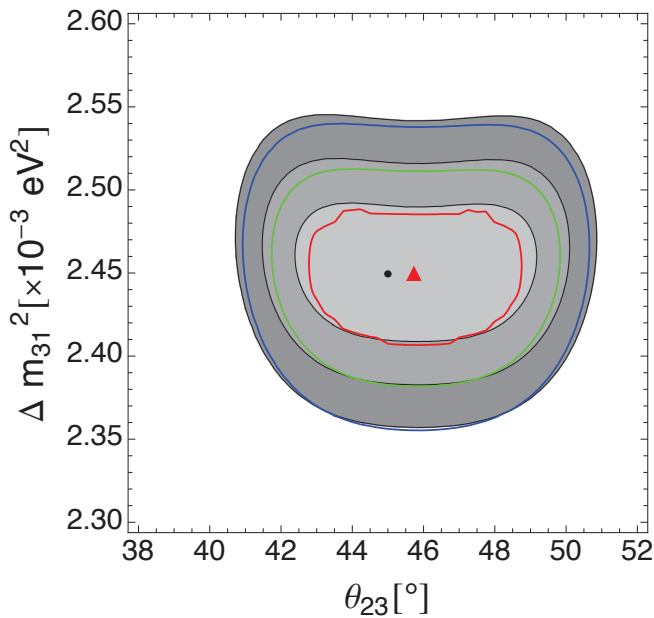}
\caption{(Color online).~Impact on the oscillation results if a different Q$^2$ selection is used to compute the true and fitted QE event rates in the oscillation analysis. In the plot is shown the result of the fit in the $\theta_{23}-\Delta m_{31}^2$ plane. The shaded area shows the confidence regions that would be obtained at 1, 2 and 3$\sigma$ if the simulated and fitted event rates were generated using the same set of migration matrices produced by GENIE $2.8.0+\nu T$ and RFGM. The colored lines show the resulting regions if the event rates are computed using matrices produced by GENIE $2.8.0+\nu T$ and RFGM are fitted with the rates computed using matrices obtained using GENIE $2.8.0$ and RFGM. The black dot show the true input value of the fit, while the red triangle shows the location of the best fit point. \label{fig:s12}}
\end{figure}

 Figure~\ref{fig:s12} shows the impact on the oscillation fit results in the case in which a different $Q^2$ selection for just the QE neutrino interaction is used to compute the true and fitted rates. In Fig.~\ref{fig:s12}  the result of the fit is represented in the $\theta_{23}-\Delta m_{31}^2$ plane. The shaded area shows the confidence regions that would be obtained at 1, 2 and 3$\sigma$ if the simulated and fitted event rates were generated using the same set of migration matrices produced by GENIE $2.8.0+\nu T$ using RFGM. The colored lines show the resulting regions if the event rates that are computed using matrices produced by GENIE $2.8.0+\nu T$ and RFGM are fitted with the rates computed using matrices obtained using GENIE $2.8.0$ and RFGM.\\
\noindent The best-fit values we found were $\theta_{23}=45.75 \ {\rm deg}$ and $\Delta m^2_{31} = 2.45\times 10^{-3} \ \textrm{eV}^2$ for a $\chi^2/{\rm ndof} = 0.78/14$. We observe a difference of 1.7\% in the fitted value for the $\theta_{23}$ mixing angle as a results of the different Q$^2$ selection between GENIE $2.8.0$ to $2.8.0+\nu T$.
\par
\begin{figure}[htbn!]
\includegraphics[width=\columnwidth]{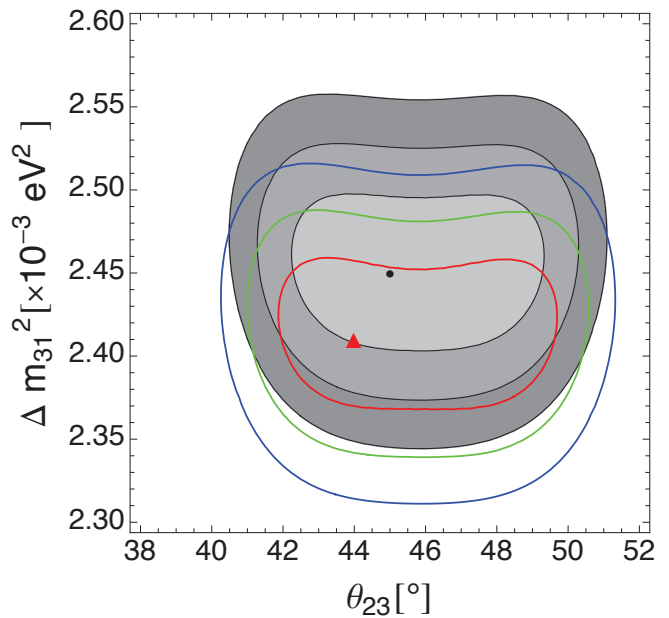}
\caption{(Color online).~Impact on the oscillation results if the spectral function nuclear model is used instead of the RFGM. In the plot is shown the result of the fit in the $\theta_{23}-\Delta m_{31}^2$ plane. The shaded area shows the confidence regions that would be obtained at 1, 2 and 3$\sigma$ if the simulated and fitted event rates are generated using the same set of migration matrices produced by GENIE $2.8.0+\nu T$ and SF. The colored lines show the resulting regions if the event rates are computed using matrices produced by GENIE $2.8.0+\nu T$ and RFGM are fitted with the rates computed using matrices obtained using GENIE $2.8.0+\nu T$ and SF. The black dot show the true input value of the fit, while the red triangle shows the location of the best fit point. \label{fig:s23}}
\end{figure}

\begin{figure}[htbn!]
\includegraphics[width=\columnwidth]{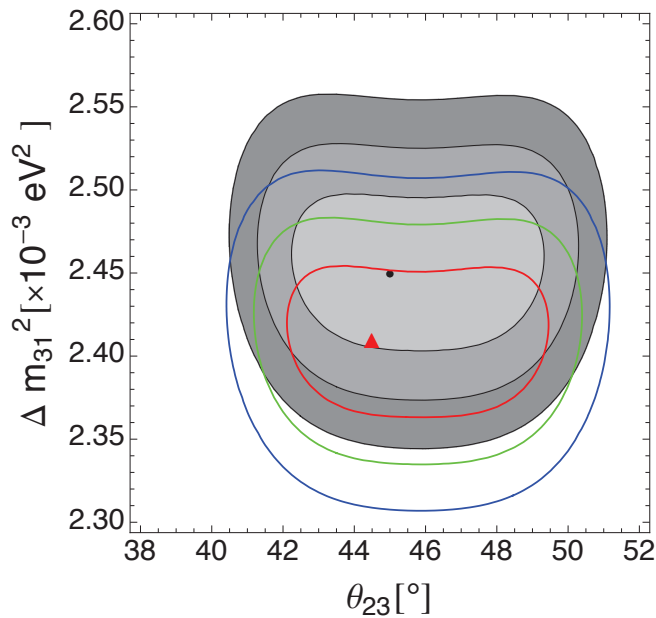}
\caption{(Color online).
~Impact on the oscillation results if the spectral function nuclear model is used instead of the RFGM and the Q$^2$ selection used in GENIE $2.8.0$. In this analysis the impact of the oscillation parameters is the convolution of a different nuclear model and different Q$^2$ selection. In the plot is shown the result of the fit in the $\theta_{23}-\Delta m_{31}^2$ plane. The shaded area shows the confidence regions that would be obtained at 1, 2 and 3$\sigma$ if the simulated and fitted event rates are generated using the same set of migration matrices produced by GENIE $2.8.0+\nu T$ and SF. The colored lines show the resulting regions if the event rates are computed using matrices produced by GENIE $2.8.0$ and RFGM are fitted with the rates computed using matrices obtained using GENIE $2.8.0+\nu T$ and SF. The black dot show the true input value of the fit, while the red triangle shows the location of the best fit point. \label{fig:s13}}
\end{figure}

 A similar analysis has been performed to pin down the difference on the fitted values of the oscillation parameters induced by the use of SF instead of RFGM as a nuclear model.
The GLoBES event distributions have been corrected using migration matrices produced by GENIE $2.8.0+\nu T$ using SF, and fitted using event distributions obtained using GENIE $2.8.0+\nu T$ and RFGM. The results are shown in Fig.~\ref{fig:s23} where the shaded area shows the confidence regions corresponding to 1, 2 and 3$\sigma$ if the simulated and fitted event rates were generated using the same set of migration matrices produced by GENIE $2.8.0+\nu T$ and SF. The colored lines show the resulting regions if the event rates that are computed using matrices produced by GENIE $2.8.0+\nu T$ and RFGM are fitted with the rates computed using matrices obtained using GENIE $2.8.0+\nu T$ and SF. \\
\noindent The results are: $\theta_{23}=44.0 \ {\rm deg}$, $\Delta m^2_{31} = 2.41\times 10^{-3} \ \textrm{eV}^2$ and $\chi^2/{\rm ndof} =  2.94/14$. The best-fit values and the confidence levels are shown in Fig.~\ref{fig:s23} with the same color code as in Fig.~\ref{fig:s12}. We observe a change in the extracted value of the oscillation parameters at the level of 2.2\% (1~$\sigma$) in the determination of the mixing angle and of 1.6\% for the mass-square splitting.
Finally we have studied the impact of different nuclear models (SF vs RFG) and of a different Q$^2$ selection on the determination of oscillation parameters we have repeated the same analysis as shown in Figs.~\ref{fig:s12} and~\ref{fig:s23}. We have corrected the GLoBES event distributions using migration matrices produced by GENIE $2.8.0+\nu T$ using SF and then we fitted them using event distributions obtained using GENIE $2.8.0$ and RFG as nuclear model.
The results are shown in Fig.~\ref{fig:s13} where the shaded area shows the confidence regions that would be obtained at 1, 2 and 3$\sigma$ if the simulated and fitted event rates  were generated using the same set of migration matrices produced by GENIE $2.8.0+\nu T$ and SF. The colored lines show the resulting regions if the event rates that are computed using matrices produced by GENIE $2.8.0$ and RFGM are fitted with the rates computed using matrices obtained using GENIE $2.8.0+\nu T$ and SF.\\
\noindent The best-fit parameters are found to be $\theta_{23}=44.5 \ {\rm deg}$ and $\Delta m^2_{31} = 2.41\times 10^{-3} \ \textrm{eV}^2$ and $\chi^2/{\rm ndof}$ is 2.94/14. In this case we found a difference of 1.1\% in the determination of the mixing angle and of 1.6\% for the mass-square splitting.
Our oscillation results are summarized in Table~\ref{tab:summary}.
It has to be noticed that the oscillation analysis discussed here is significantly simplified, in that we have not used the near detector to constrain the model. In a more sophisticated experimental analysis, particularly when there is no consistency in the modeling of different reaction channels in the neutrino interaction generator, it would be natural to reweight them to get a better description of data at the near detector. Such a procedure would reduce somehow the impact of the QE model on the oscillation parameters. However,  more complicated problems, like propagating uncertainties on quantities measured at the near detector with a different beam energy and angles, would need to be propagated to the far detector.
\begin{table}[htbn!]
\begin{center}
\caption{Summary of the main impact on the oscillation parameters for the different scenarios studied in this work. The true values for the disappearance oscillation parameters are $\theta_{23}=45~deg$ and $\Delta m^2_{31}=2.45\times10^{-3}\,\mathrm{eV}^2$. The number of degrees of freedom in the fit is $n-p=14$, where $n$ is the number of energy bins and $p$ is the number of oscillation parameters that are being estimated from the fit. }
\begin{tabular}{l@{\quad}l|@{\quad}c@{\quad}c}
True & Fitted & $\theta_{23,min} $ & $\Delta
m^2_{31,min}[\textrm{eV}^2] $ \\ \hline
RFGM$_{2.8.0+\nu T}$ & RFGM$_{2.8.0}$    & 45.7~deg & 2.45$\times 10^{-3}$ \\ \hline
SF$_{2.8.0+ \nu T}$    & RFGM$_{2.8.0+ \nu T}$  & 44~deg & 2.41$\times 10^{-3}$  \\ \hline
SF$_{2.8.0+ \nu T}$    & RFGM$_{2.8.0}$    & 44.5~deg & 2.41$\times 10^{-3}$ \\ \hline
\end{tabular}
\label{tab:summary}
\end{center}
\end{table}
%
%
%
%
%
%
\section{Conclusions}
\label{sec:conclusions}
%
We have implemented the description of the nuclear ground state based on realistic spectral functions --
widely and successfully employed in the analysis of electron-nucleus scattering data -- into
the GENIE generator of neutrino interactions. We hope that this work will be soon included in an official GENIE release.
Compared to the RFGM, the spectral function approach predicts the occurrence of nucleons carrying momenta
much larger that the Fermi momentum, and high removal energy, in the target ground state. Energy and momentum conservation implies
a strong correlation between high momentum and high removal energy. As a consequence, knock out of a high momentum
nucleon leaves the residual system with high excitation energy. The nuclear final state of these processes is a two particle-two hole
state, as one of the spectator particles is excited to the continuum.
Besides introducing a more realistic model of nuclear dynamics, we have improved the simulation of the kinematics variables
of the outgoing particles, requiring that the momentum and energy transfer entering the definition of the selected $Q^2$ satisfy
the requirements of energy and momentum conservation. In this context, it has to be emphasized that, as the nuclear response to
electroweak interactions is function of {\em two} variables, e.g. the momentum and energy transfer $|{\bf q}|$ and $\omega$,
the simulation algorithm based on $Q^2$ selection may not be the most effective.
The implementation of the spectral functions and the improved $Q^2$ selection have been validated through comparison
to electron scattering data for different targets and kinematical setups. The simulated cross sections also agree with the
results of theoretical calculations based on the same dynamical model. We note that the large body of precise electron scattering data
should be exploited to perform similar comparisons using all existing neutrino event generators.
The neutrino interaction events generated with the modified GENIE, that we refer to as GENIE $2.8.0+\nu T$, have been studied as
a function of both $Q^2$ and the observed kinematical variables of the outgoing charged lepton. In all instances, the new features
introduced in the simulation process turn out to have sizable effects.
\begin{figure}[t!]
\includegraphics[width=0.9\columnwidth]{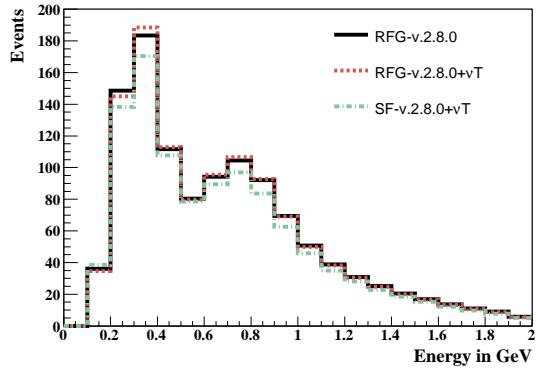}\label{fig:qe_s_all}
\caption{(Color online).~QE only event distributions as a function of the reconstructed neutrino energy for the different nuclear models and Q$^2$ selection. The oscillation parameters have been set to their values in Eq.~\eqref{eq:oscparams}, and we have included also detection efficiencies. The black solid line shows the spectrum from the RFGM of GENIE $2.8.0$, red dotted line shows the spectrum from GENIE $2.8.0+\nu T$ and green dotted-dashed line represents the signal distribution from the SF code of GENIE $2.8.0+\nu T$.}
\label{fig:ev2}
\end{figure}
\begin{figure*}[htbn!]
\subfigure[~Resonance]{\includegraphics[width=0.9\columnwidth, height=0.74\columnwidth]{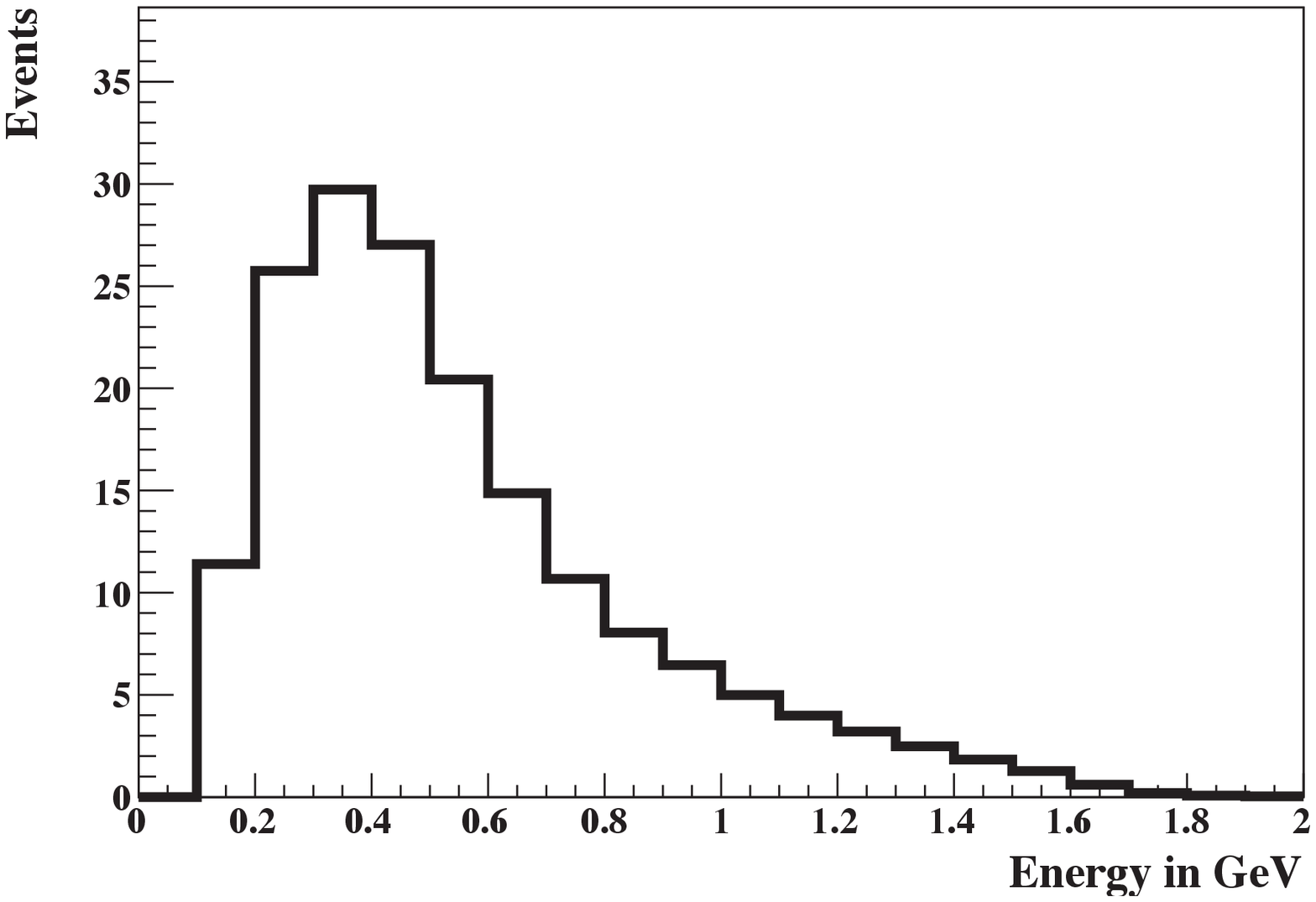}\label{fig:s_res}}
\subfigure[~Non-resonance]{\includegraphics[width=0.9\columnwidth, height=0.74\columnwidth]{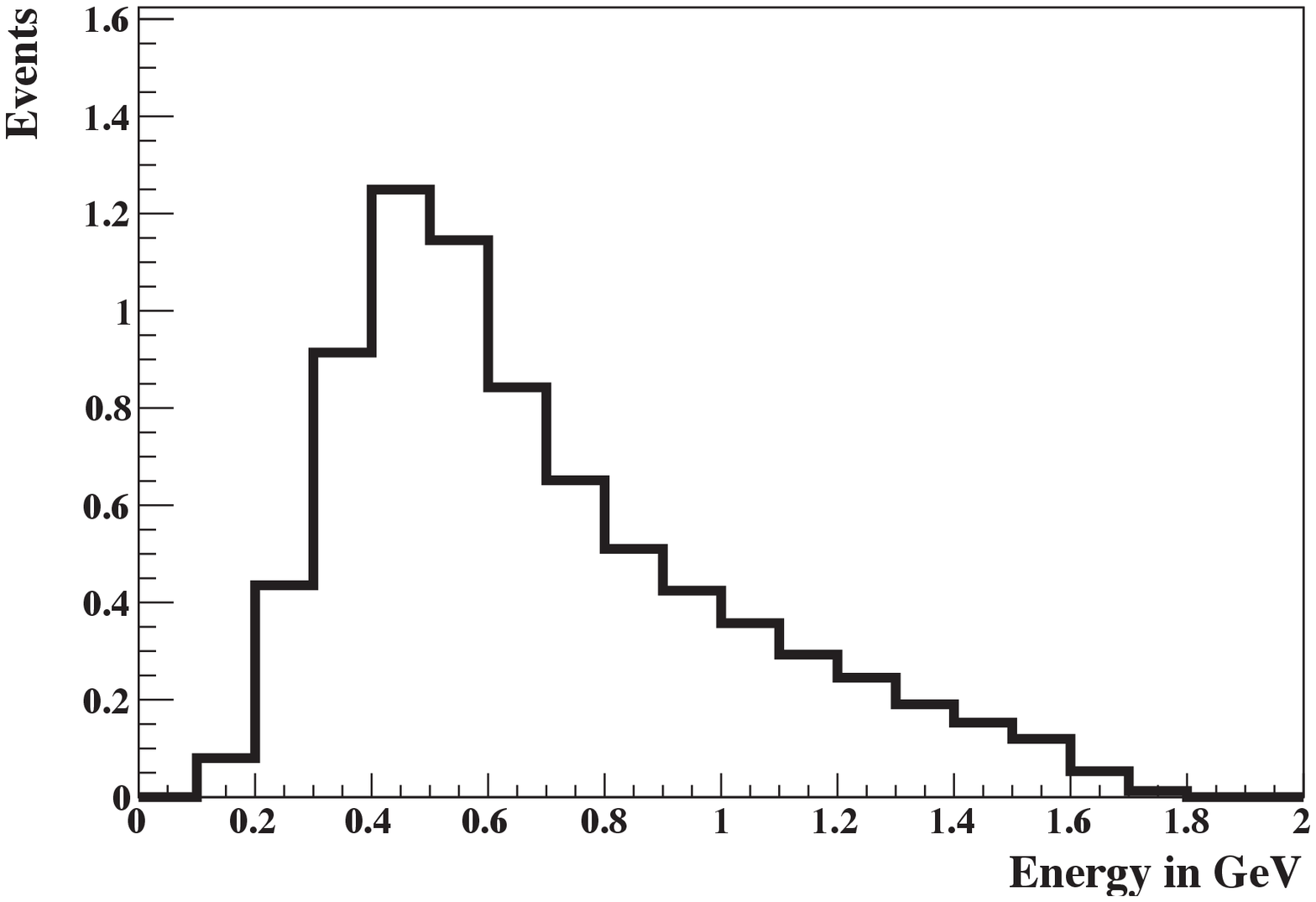}\label{fig:s_nres}}
\subfigure[~MEC/2p2h]{\includegraphics[width=0.9\columnwidth, height=0.74\columnwidth]{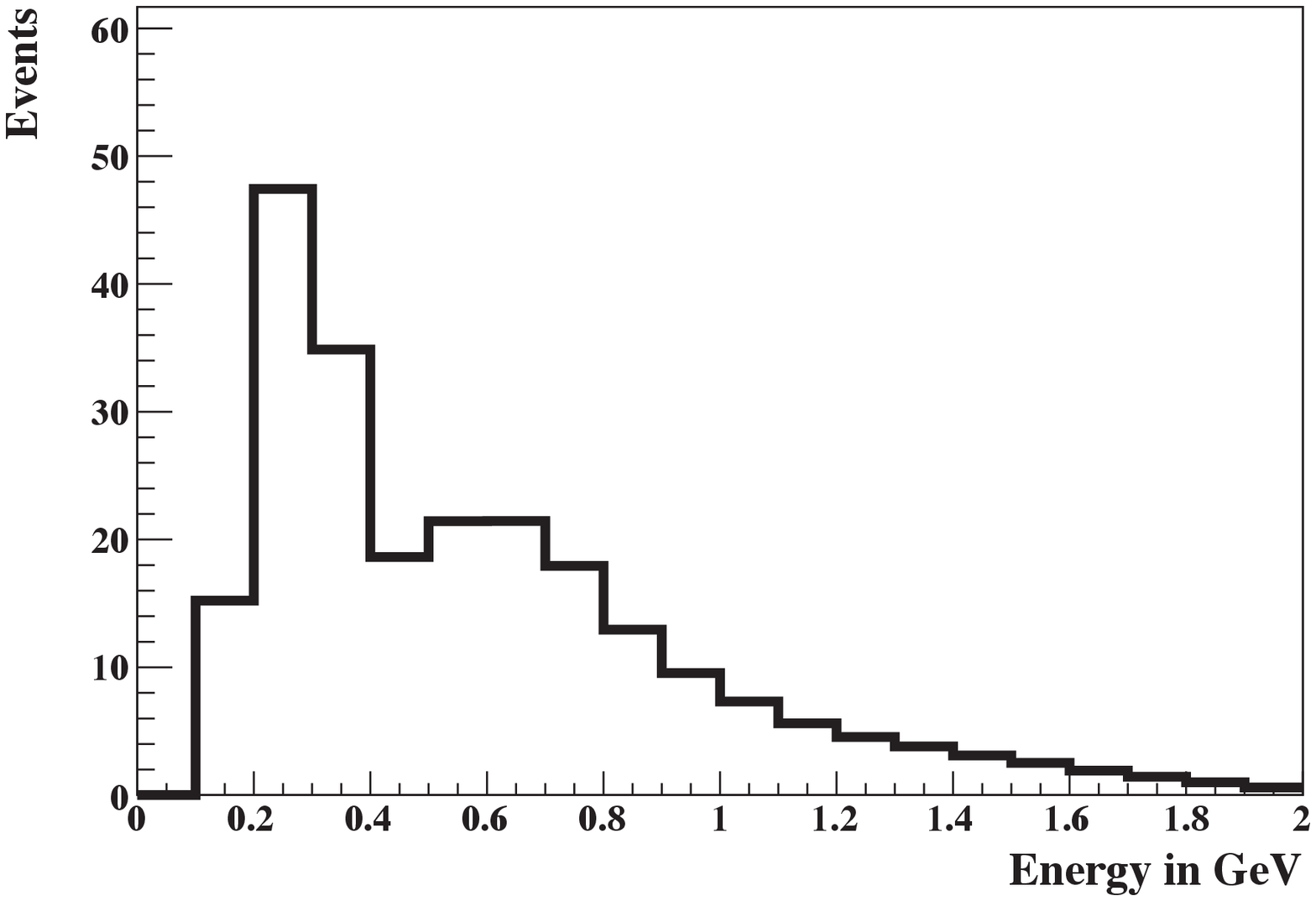}\label{fig:s_mec}}
\caption{Event distributions as a function of the reconstructed neutrino energy for the different nuclear models and Q$^2$ selection. The neutrino energy has been reconstructed assuming a pure QE events as in Eq.~\eqref{eq:reconstructed_enu}. The oscillation parameters have been set to their values in Eq.~\eqref{eq:oscparams} and we have included also detection efficiencies. }
\label{fig:ev1}
\end{figure*}
\begin{figure*}[htbn!]
\subfigure[~QE-RFGM, GENIE $2.8.0$]{\includegraphics[width=0.9\columnwidth, height=0.7\columnwidth]{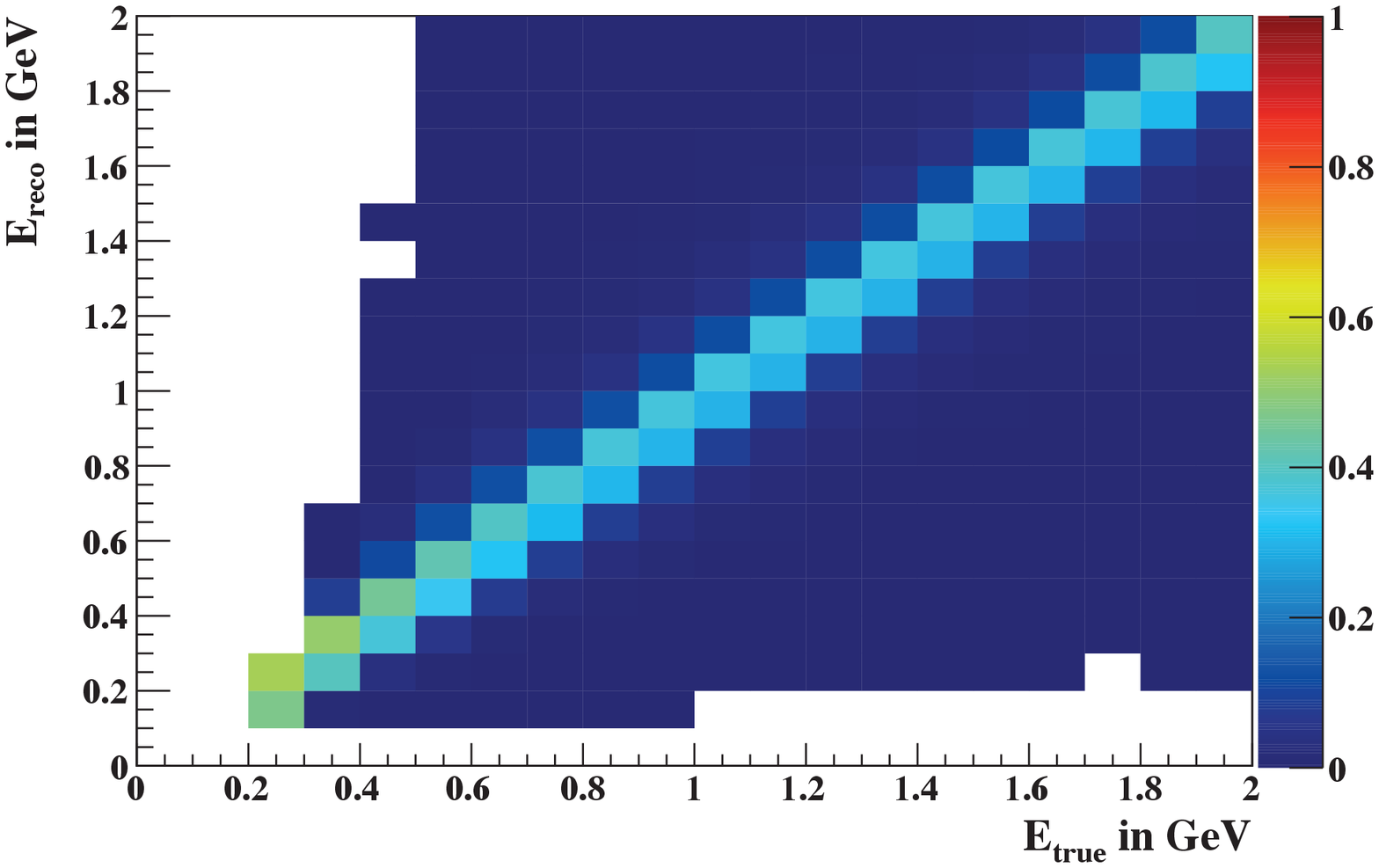}\label{fig:mm_0}}
\subfigure[~QE-RFGM, GENIE $2.8.0+\nu T$]{\includegraphics[width=0.9\columnwidth, height=0.7\columnwidth]{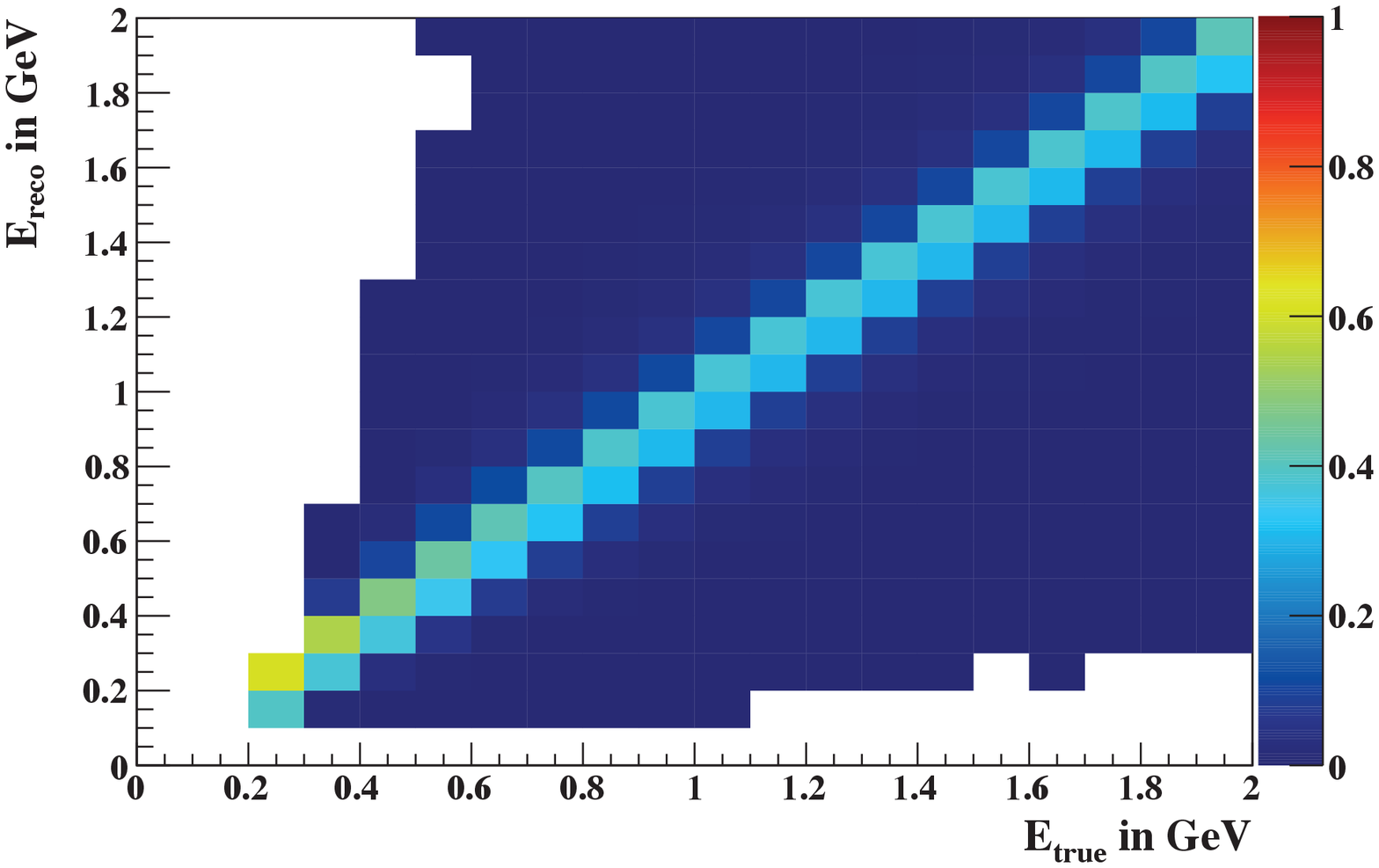}\label{fig:mm_plus}}
\subfigure[~QE-SF model, GENIE $2.8.0+\nu T$]{\includegraphics[width=0.9\columnwidth, height=0.7\columnwidth]{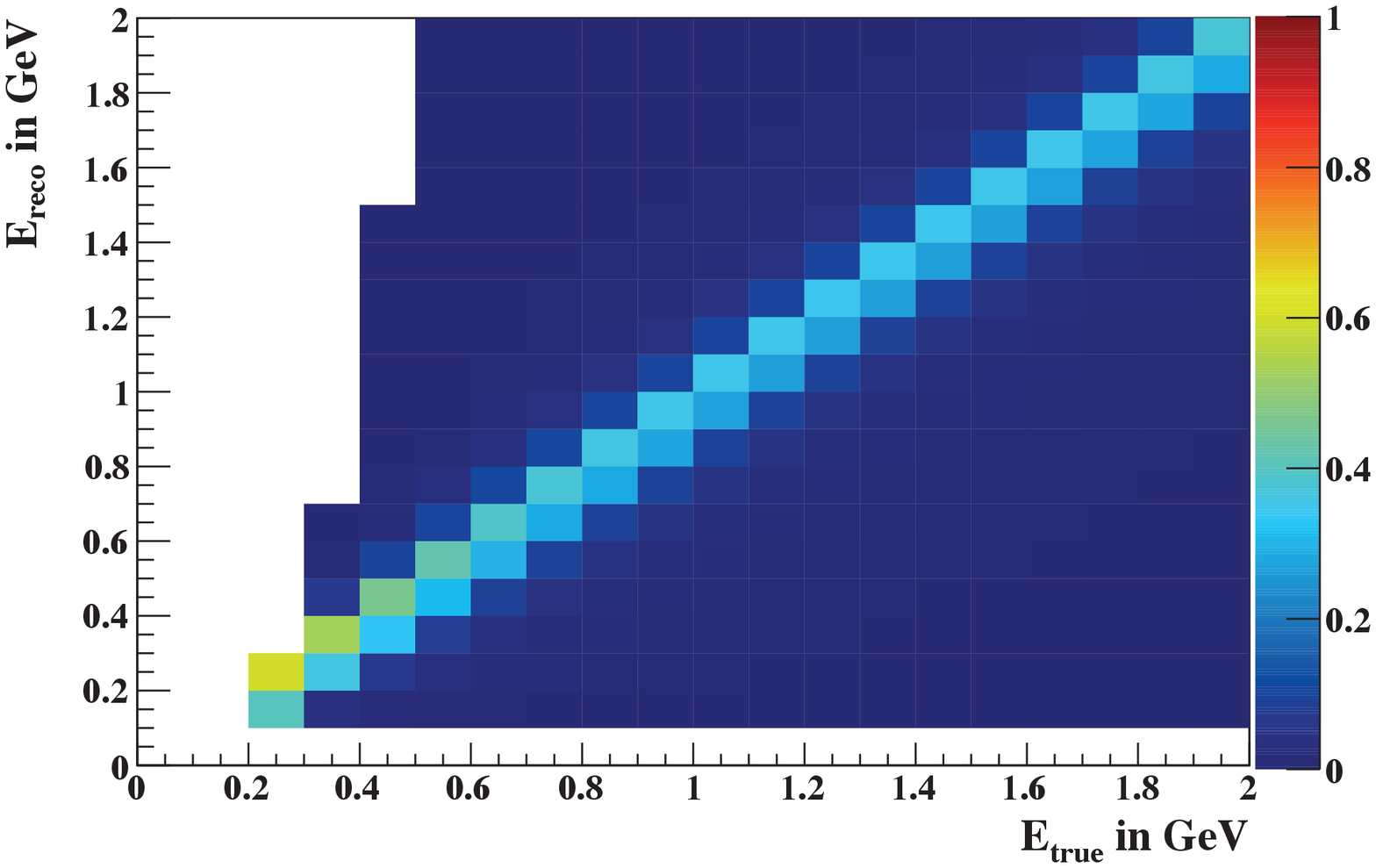}\label{fig:mm_sf}}
\subfigure[~Resonant production]{\includegraphics[width=0.9\columnwidth, height=0.7\columnwidth]{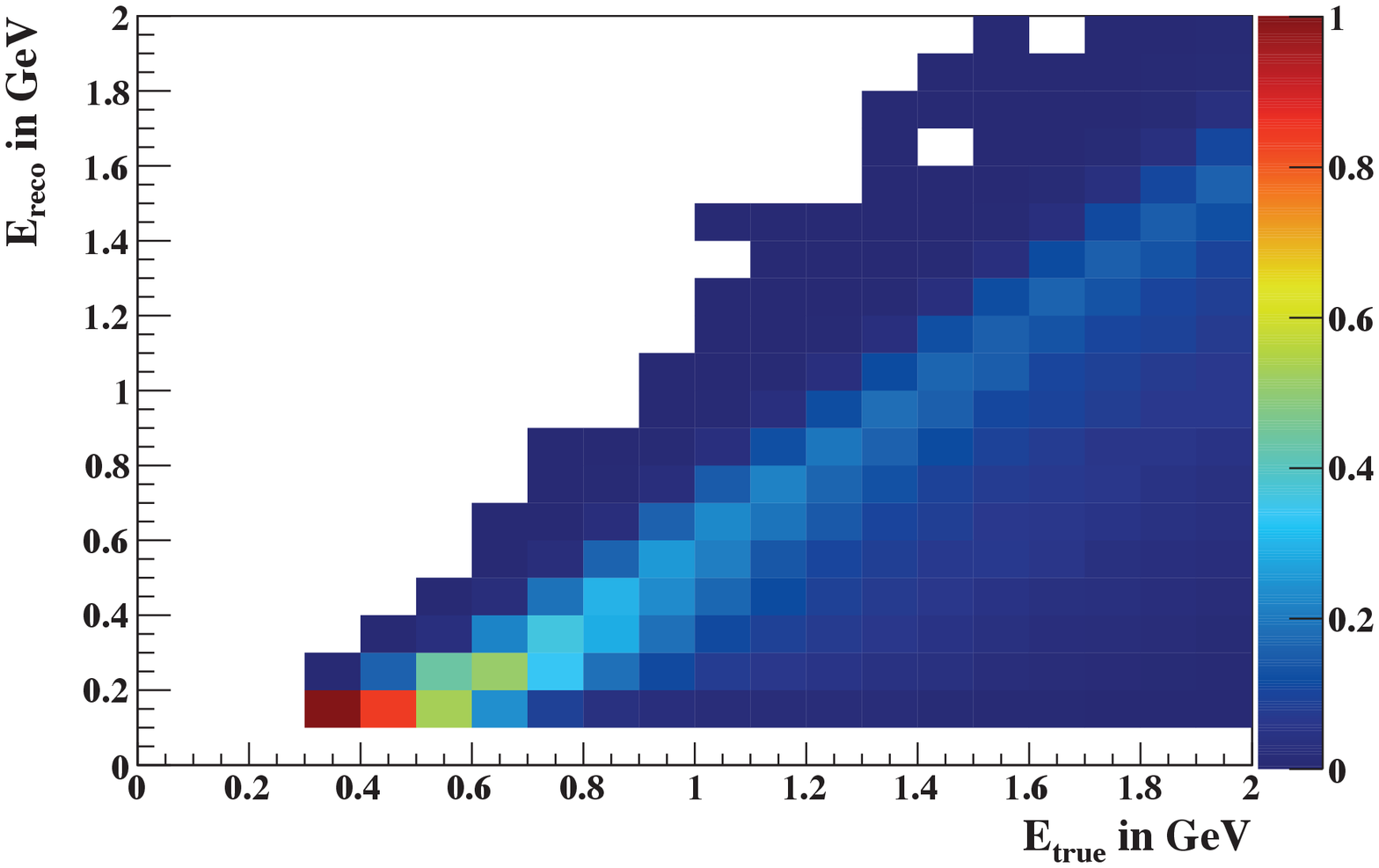}\label{fig:mm_res}}
\subfigure[~Non-resonant production]{\includegraphics[width=0.9\columnwidth, height=0.7\columnwidth]{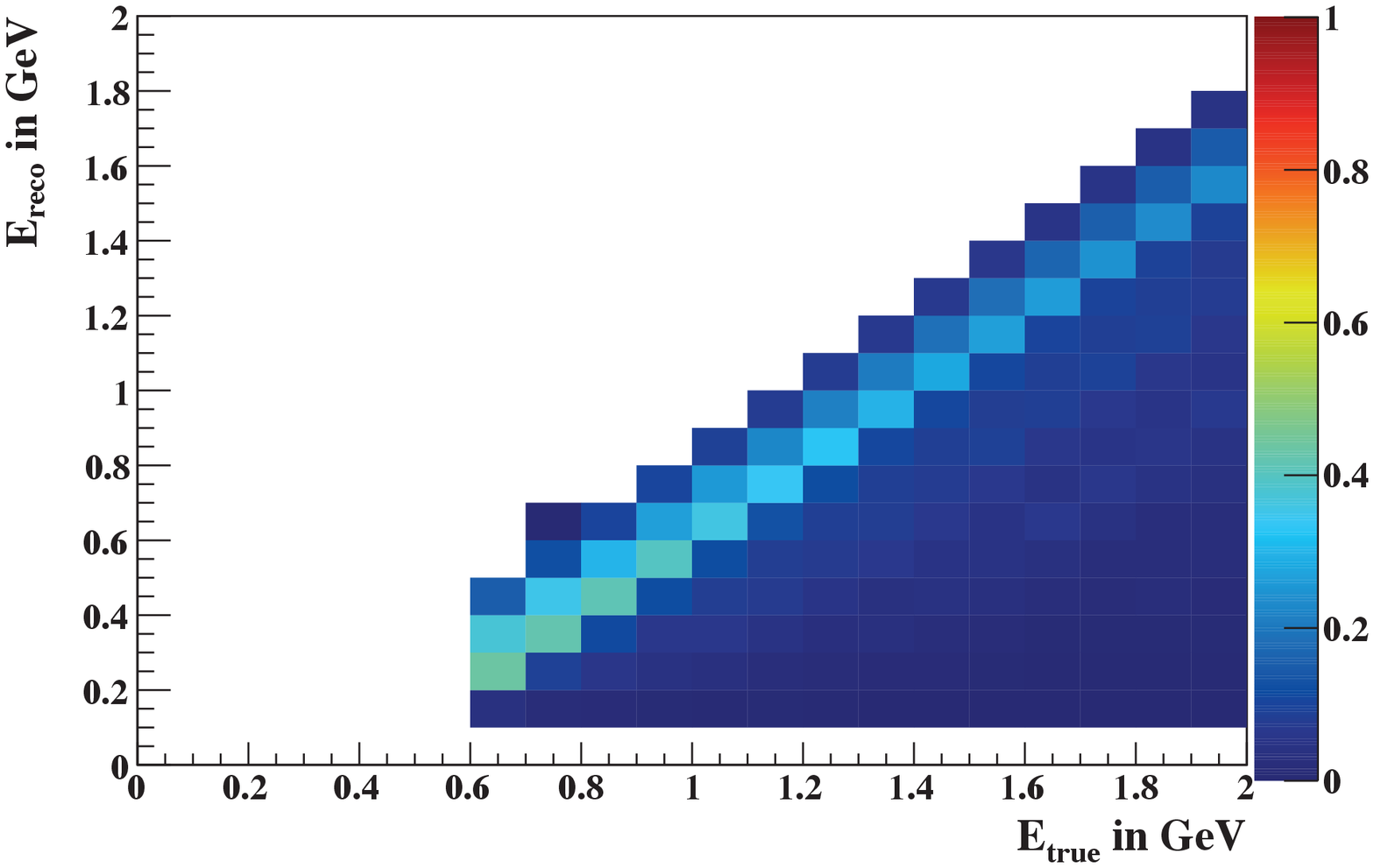}\label{fig:mm_nres}}
\subfigure[~MEC/2p2h interactions]{\includegraphics[width=0.9\columnwidth, height=0.7\columnwidth]{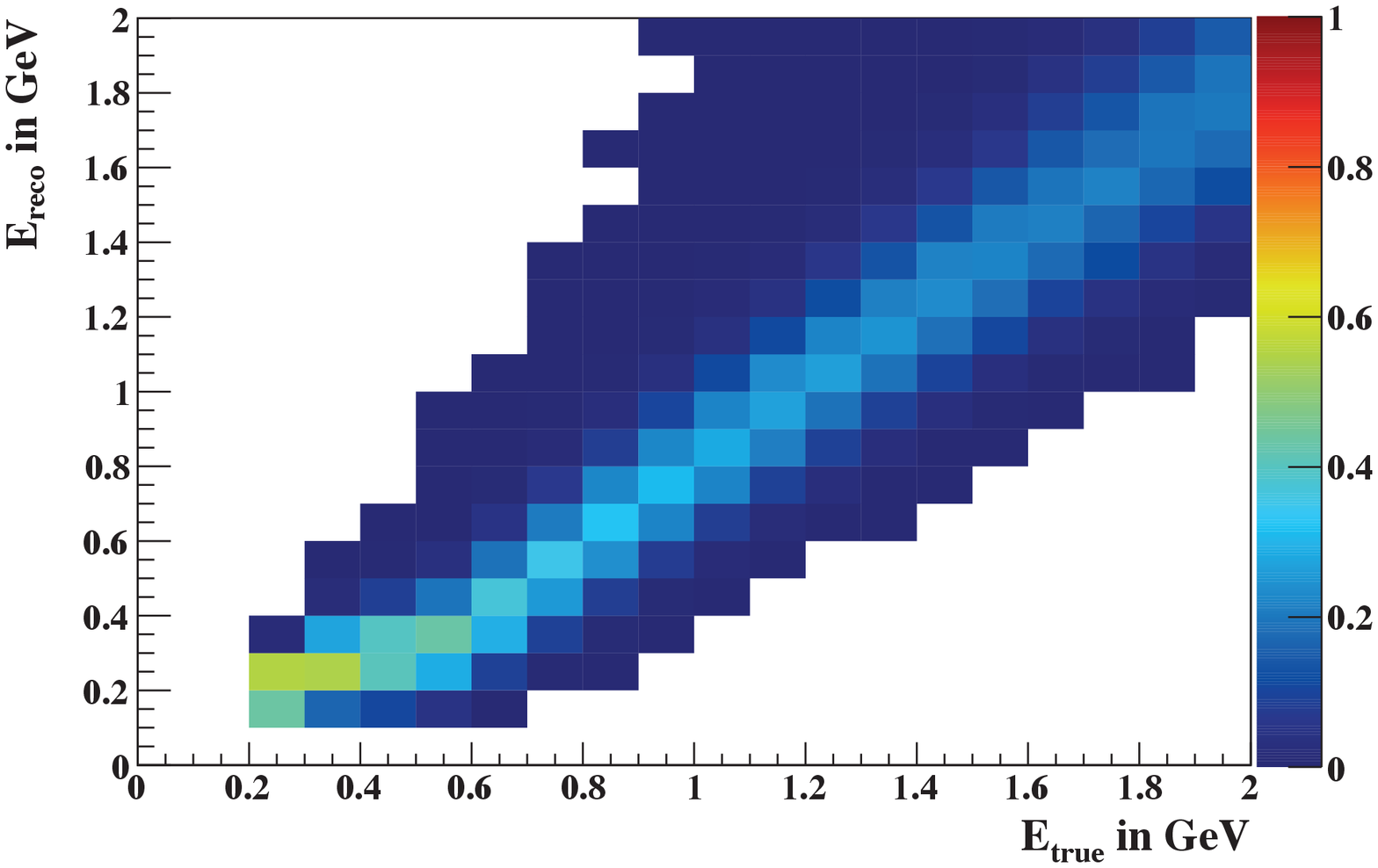}\label{fig:mm_mec}}
\caption{(Color online).~Migration matrices $M_{ij}$ for the three QE and QE-like channels: QE-RFGM of GENIE $2.8.0$ (a), QE-RFGM with the new Q$^2$ selection of GENIE $2.8.0+\nu T$ (b), SF of GENIE $2.8.0+\nu T$ (c), resonant production (d), non-resonant production (e) and MEC/2p2h (f). The GENIE $2.8.0$ generator was used to produce the $M_{ij}$ with 200,000 neutrino interactions for each of the true energy bin.}
\label{fig:mat1}
\end{figure*}
The oscillation analysis carried out using the GLoBES sensitivity framework indicates that the treatment of the nuclear
ground state and the $Q^2$ selection do have a {\em non-negligible} influence on the determination of the atmospheric oscillation
parameter in a typical $\nu_\mu$ disappearance experiment. Note that, while our study focused on
CCQE interactions only,  the replacement of the RGFM with the spectral function approach, implying an improved
treatment of the initial nuclear state, affects {\em all} reaction channels, including resonance production and DIS.
The modification of the oscillation parameters resulting from our calculations must be regarded as an indication of what
is the effect of nuclear models in a very simple oscillation analysis. Neutrino experiments use much more complicated analysis and
they especially use to reweight the simulation provided by the event generator to get a better description of data at the near detector. Such a procedure would lead to a reduction of
the impact of the QE model on the oscillation parameters, but would rise questions as to how the systematic uncertainties are propagated from the near to the far detector, where the beam is different and the phase space of neutrino interactions is different as well. Our analysis is very simple and provides a way to evaluate the effect of nuclear modeling using the near detector just for the normalization of the integrated inclusive cross section.
In summary, the new implementation of the spectral functions, based on
a more efficient sampling algorithm and improved $Q^2$ selection, represents a
step forward towards the understanding and characterization of neutrino
interactions. As far as processes driven by one-nucleon currents are
concerned, the CCQE sector is now described within a consistent framework,
in which energy and momentum conservation, providing the link between
initial and final state kinematics, is properly taken into account. The
numerical results obtained from GENIE $2.8.0+\nu T$ turn out to be in
agreement with electron scattering data collected in different kinematical
setups and using different nuclear targets.
\section*{Acknowledgements}
We are deeply indebted to S. Dytman and H. Gallagher for their support on the GENIE implementation and many illuminating discussions. We are grateful to N. Rocco for providing Fig.~\ref{F00}, and to P. Coloma and P. Huber for their advice and support on the oscillation analysis. Finally, we are pleased to acknowledge the support of the Department of Physics and Center for Neutrino Physics of Virginia Tech, that made our work possible. This work has been supported by the National Science Foundation under award number PHY-1352106. A. M. A. was supported by JSPS under Grant No. PE 13056.
\appendix
\section{Migration matrices and event distributions\label{app:matrices} }
Figure~\ref{fig:ev2} shows the pure QE event distributions as a function of the reconstructed neutrino energy for the different nuclear models and Q$^2$ selection.
The oscillation parameters have been set to their values in Eq.~\eqref{eq:oscparams} and we have included also detection efficiencies. The distributions for QE-like events are shown in Fig.~\ref{fig:ev1}.
In all cases, the reconstructed neutrino energy is calculated assuming a pure QE event and according to Eq.~\eqref{eq:reconstructed_enu}.
Figure~\ref{fig:mat1} shows the migration matrices for the QE and QE-like for neutrino interactions on $^{12}$C.
The three QE matrices (\ref{fig:mm_0},~\ref{fig:mm_plus} and~\ref{fig:mm_sf}) were computed using GENIE $2.8.0$ and GENIE $2.8.0+\nu T$ with RFGM and SF as nuclear models.
The QE-like matrices (\ref{fig:mm_res},~\ref{fig:mm_nres} and~\ref{fig:mm_mec}) were computed using the GENIE $2.8.0$ official release and they are common to the three models investigated in this publication.
%
\bibliographystyle{apsrev}
\bibliography{references}

\begin{thebibliography}{39}
\expandafter\ifx\csname natexlab\endcsname\relax\def\natexlab#1{#1}\fi
\expandafter\ifx\csname bibnamefont\endcsname\relax
  \def\bibnamefont#1{#1}\fi
\expandafter\ifx\csname bibfnamefont\endcsname\relax
  \def\bibfnamefont#1{#1}\fi
\expandafter\ifx\csname citenamefont\endcsname\relax
  \def\citenamefont#1{#1}\fi
\expandafter\ifx\csname url\endcsname\relax
  \def\url#1{\texttt{#1}}\fi
\expandafter\ifx\csname urlprefix\endcsname\relax\def\urlprefix{URL }\fi
\providecommand{\bibinfo}[2]{#2}
\providecommand{\eprint}[2][]{\url{#2}}

\bibitem[{\citenamefont{Abe et~al.}(2012)}]{Abe:2011fz}
\bibinfo{author}{\bibfnamefont{Y.}~\bibnamefont{Abe}} \bibnamefont{et~al.}
  (\bibinfo{collaboration}{DOUBLE-CHOOZ Collaboration}),
  \bibinfo{journal}{Phys.Rev.Lett.} \textbf{\bibinfo{volume}{108}},
  \bibinfo{pages}{131801} (\bibinfo{year}{2012}), \eprint{1112.6353}.

\bibitem[{\citenamefont{An et~al.}(2014)}]{An:2013zwz}
\bibinfo{author}{\bibfnamefont{F.}~\bibnamefont{An}} \bibnamefont{et~al.}
  (\bibinfo{collaboration}{Daya Bay Collaboration}),
  \bibinfo{journal}{Phys.Rev.Lett.} \textbf{\bibinfo{volume}{112}},
  \bibinfo{pages}{061801} (\bibinfo{year}{2014}), \eprint{1310.6732}.

\bibitem[{\citenamefont{Ahn et~al.}(2012)}]{Ahn:2012nd}
\bibinfo{author}{\bibfnamefont{J.}~\bibnamefont{Ahn}} \bibnamefont{et~al.}
  (\bibinfo{collaboration}{RENO collaboration}),
  \bibinfo{journal}{Phys.Rev.Lett.} \textbf{\bibinfo{volume}{108}},
  \bibinfo{pages}{191802} (\bibinfo{year}{2012}), \eprint{1204.0626}.

\bibitem[{\citenamefont{Abe et~al.}(2014)}]{Abe:2013hdq}
\bibinfo{author}{\bibfnamefont{K.}~\bibnamefont{Abe}} \bibnamefont{et~al.}
  (\bibinfo{collaboration}{T2K Collaboration}),
  \bibinfo{journal}{Phys.Rev.Lett.} \textbf{\bibinfo{volume}{112}},
  \bibinfo{pages}{061802} (\bibinfo{year}{2014}), \eprint{1311.4750}.

\bibitem[{\citenamefont{Abe et~al.}(2013)}]{Abe:2013xua}
\bibinfo{author}{\bibfnamefont{K.}~\bibnamefont{Abe}} \bibnamefont{et~al.}
  (\bibinfo{collaboration}{T2K Collaboration}), \bibinfo{journal}{Phys.Rev.}
  \textbf{\bibinfo{volume}{D88}}, \bibinfo{pages}{032002}
  (\bibinfo{year}{2013}), \eprint{1304.0841}.

\bibitem[{\citenamefont{Aguilar-Arevalo
  et~al.}(2010{\natexlab{a}})}]{miniboone_ccqe}
\bibinfo{author}{\bibfnamefont{A.}~\bibnamefont{Aguilar-Arevalo}}
  \bibnamefont{et~al.} (\bibinfo{collaboration}{MiniBooNE Collaboration}),
  \bibinfo{journal}{Phys.Rev.} \textbf{\bibinfo{volume}{D81}},
  \bibinfo{pages}{092005} (\bibinfo{year}{2010}{\natexlab{a}}),
  \eprint{1002.2680}.

\bibitem[{\citenamefont{Aguilar-Arevalo
  et~al.}(2010{\natexlab{b}})}]{miniboone_nc}
\bibinfo{author}{\bibfnamefont{A.}~\bibnamefont{Aguilar-Arevalo}}
  \bibnamefont{et~al.} (\bibinfo{collaboration}{MiniBooNE Collaboration}),
  \bibinfo{journal}{Phys.Rev.} \textbf{\bibinfo{volume}{D82}},
  \bibinfo{pages}{092005} (\bibinfo{year}{2010}{\natexlab{b}}).

\bibitem[{\citenamefont{R. et~al.}(2006)}]{K2K_ccqe}
\bibinfo{author}{\bibfnamefont{G.}~\bibnamefont{R.}} \bibnamefont{et~al.}
  (\bibinfo{collaboration}{K2K Collaboration}), \bibinfo{journal}{Phys.Rev.}
  \textbf{\bibinfo{volume}{D74}}, \bibinfo{pages}{052002}
  (\bibinfo{year}{2006}).

\bibitem[{\citenamefont{Benhar et~al.}(2005)\citenamefont{Benhar, Farina,
  Nakamura, Sakuda, and Seki}}]{PhysRevD.72.053005}
\bibinfo{author}{\bibfnamefont{O.}~\bibnamefont{Benhar}},
  \bibinfo{author}{\bibfnamefont{N.}~\bibnamefont{Farina}},
  \bibinfo{author}{\bibfnamefont{H.}~\bibnamefont{Nakamura}},
  \bibinfo{author}{\bibfnamefont{M.}~\bibnamefont{Sakuda}}, \bibnamefont{and}
  \bibinfo{author}{\bibfnamefont{R.}~\bibnamefont{Seki}},
  \bibinfo{journal}{Phys. Rev. D} \textbf{\bibinfo{volume}{72}},
  \bibinfo{pages}{053005} (\bibinfo{year}{2005}).

\bibitem[{\citenamefont{Benhar and Meloni}(2007)}]{Benhar:2006nr}
\bibinfo{author}{\bibfnamefont{O.}~\bibnamefont{Benhar}} \bibnamefont{and}
  \bibinfo{author}{\bibfnamefont{D.}~\bibnamefont{Meloni}},
  \bibinfo{journal}{Nucl.Phys.} \textbf{\bibinfo{volume}{A789}},
  \bibinfo{pages}{379} (\bibinfo{year}{2007}), \eprint{hep-ph/0610403}.

\bibitem[{\citenamefont{Martini et~al.}(2012)\citenamefont{Martini, Ericson,
  and Chanfray}}]{Martini:2012fa}
\bibinfo{author}{\bibfnamefont{M.}~\bibnamefont{Martini}},
  \bibinfo{author}{\bibfnamefont{M.}~\bibnamefont{Ericson}}, \bibnamefont{and}
  \bibinfo{author}{\bibfnamefont{G.}~\bibnamefont{Chanfray}},
  \bibinfo{journal}{Phys.Rev.} \textbf{\bibinfo{volume}{D85}},
  \bibinfo{pages}{093012} (\bibinfo{year}{2012}), \eprint{1202.4745}.

\bibitem[{\citenamefont{Meloni and Martini}(2012)}]{Meloni:2012fq}
\bibinfo{author}{\bibfnamefont{D.}~\bibnamefont{Meloni}} \bibnamefont{and}
  \bibinfo{author}{\bibfnamefont{M.}~\bibnamefont{Martini}},
  \bibinfo{journal}{Phys.Lett.} \textbf{\bibinfo{volume}{B716}},
  \bibinfo{pages}{186} (\bibinfo{year}{2012}), \eprint{1203.3335}.

\bibitem[{\citenamefont{Nieves et~al.}(2012)\citenamefont{Nieves, S\'anchez,
  Simo, and Vacas}}]{Nieves:2012yz}
\bibinfo{author}{\bibfnamefont{J.}~\bibnamefont{Nieves}},
  \bibinfo{author}{\bibfnamefont{F.}~\bibnamefont{S\'anchez}},
  \bibinfo{author}{\bibfnamefont{I.~R.} \bibnamefont{Simo}}, \bibnamefont{and}
  \bibinfo{author}{\bibfnamefont{M.~J.~V.} \bibnamefont{Vacas}},
  \bibinfo{journal}{Phys. Rev. D} \textbf{\bibinfo{volume}{85}},
  \bibinfo{pages}{113008} (\bibinfo{year}{2012}).

\bibitem[{\citenamefont{Lalakulich et~al.}(2012)\citenamefont{Lalakulich,
  Mosel, and Gallmeister}}]{Lalakulich:2012hs}
\bibinfo{author}{\bibfnamefont{O.}~\bibnamefont{Lalakulich}},
  \bibinfo{author}{\bibfnamefont{U.}~\bibnamefont{Mosel}}, \bibnamefont{and}
  \bibinfo{author}{\bibfnamefont{K.}~\bibnamefont{Gallmeister}},
  \bibinfo{journal}{Phys.Rev.} \textbf{\bibinfo{volume}{C86}},
  \bibinfo{pages}{054606} (\bibinfo{year}{2012}), \eprint{1208.3678}.

\bibitem[{\citenamefont{Martini et~al.}(2013)\citenamefont{Martini, Ericson,
  and Chanfray}}]{Martini:2012uc}
\bibinfo{author}{\bibfnamefont{M.}~\bibnamefont{Martini}},
  \bibinfo{author}{\bibfnamefont{M.}~\bibnamefont{Ericson}}, \bibnamefont{and}
  \bibinfo{author}{\bibfnamefont{G.}~\bibnamefont{Chanfray}},
  \bibinfo{journal}{Phys.Rev.} \textbf{\bibinfo{volume}{D87}},
  \bibinfo{pages}{013009} (\bibinfo{year}{2013}), \eprint{1211.1523}.

\bibitem[{\citenamefont{Mosel et~al.}(2013)\citenamefont{Mosel, Lalakulich, and
  Gallmeister}}]{Mosel:2013fxa}
\bibinfo{author}{\bibfnamefont{U.}~\bibnamefont{Mosel}},
  \bibinfo{author}{\bibfnamefont{O.}~\bibnamefont{Lalakulich}},
  \bibnamefont{and}
  \bibinfo{author}{\bibfnamefont{K.}~\bibnamefont{Gallmeister}}
  (\bibinfo{year}{2013}), \eprint{1311.7288}.

\bibitem[{\citenamefont{Benhar et~al.}(1993)\citenamefont{Benhar,
  Pandharipande, and Pieper}}]{RevModPhys.65.817}
\bibinfo{author}{\bibfnamefont{O.}~\bibnamefont{Benhar}},
  \bibinfo{author}{\bibfnamefont{V.~R.} \bibnamefont{Pandharipande}},
  \bibnamefont{and} \bibinfo{author}{\bibfnamefont{S.~C.}
  \bibnamefont{Pieper}}, \bibinfo{journal}{Rev. Mod. Phys.}
  \textbf{\bibinfo{volume}{65}}, \bibinfo{pages}{817} (\bibinfo{year}{1993}),
  \urlprefix\url{http://link.aps.org/doi/10.1103/RevModPhys.65.817}.

\bibitem[{\citenamefont{Pandharipande et~al.}(1997)\citenamefont{Pandharipande,
  Sick, and Huberts}}]{RevModPhys.69.981}
\bibinfo{author}{\bibfnamefont{V.~R.} \bibnamefont{Pandharipande}},
  \bibinfo{author}{\bibfnamefont{I.}~\bibnamefont{Sick}}, \bibnamefont{and}
  \bibinfo{author}{\bibfnamefont{P.~K. A.~d.} \bibnamefont{Huberts}},
  \bibinfo{journal}{Rev. Mod. Phys.} \textbf{\bibinfo{volume}{69}},
  \bibinfo{pages}{981} (\bibinfo{year}{1997}),
  \urlprefix\url{http://link.aps.org/doi/10.1103/RevModPhys.69.981}.

\bibitem[{\citenamefont{Benhar et~al.}(1994)\citenamefont{Benhar, Fabrocini,
  Fantoni, and Sick}}]{LDA}
\bibinfo{author}{\bibfnamefont{O.}~\bibnamefont{Benhar}},
  \bibinfo{author}{\bibfnamefont{A.}~\bibnamefont{Fabrocini}},
  \bibinfo{author}{\bibfnamefont{S.}~\bibnamefont{Fantoni}}, \bibnamefont{and}
  \bibinfo{author}{\bibfnamefont{I.}~\bibnamefont{Sick}},
  \bibinfo{journal}{Nuclear Physics A} \textbf{\bibinfo{volume}{579}},
  \bibinfo{pages}{493 } (\bibinfo{year}{1994}), ISSN \bibinfo{issn}{0375-9474}.

\bibitem[{\citenamefont{Ankowski and Sobczyk}(2008)}]{Ankowski:2007uy}
\bibinfo{author}{\bibfnamefont{A.~M.} \bibnamefont{Ankowski}} \bibnamefont{and}
  \bibinfo{author}{\bibfnamefont{J.~T.} \bibnamefont{Sobczyk}},
  \bibinfo{journal}{Phys.Rev.} \textbf{\bibinfo{volume}{C77}},
  \bibinfo{pages}{044311} (\bibinfo{year}{2008}), \eprint{0711.2031}.

\bibitem[{\citenamefont{Benhar et~al.}(2008)\citenamefont{Benhar, Day, and
  Sick}}]{Benhar:2006wy}
\bibinfo{author}{\bibfnamefont{O.}~\bibnamefont{Benhar}},
  \bibinfo{author}{\bibfnamefont{D.}~\bibnamefont{Day}}, \bibnamefont{and}
  \bibinfo{author}{\bibfnamefont{I.}~\bibnamefont{Sick}},
  \bibinfo{journal}{Rev.Mod.Phys.} \textbf{\bibinfo{volume}{80}},
  \bibinfo{pages}{189} (\bibinfo{year}{2008}), \eprint{nucl-ex/0603029}.

\bibitem[{\citenamefont{de~Forest~Jr.}(1983)}]{Forest83}
\bibinfo{author}{\bibfnamefont{T.}~\bibnamefont{de~Forest~Jr.}},
  \bibinfo{journal}{Nucl. Phys.} \textbf{\bibinfo{volume}{A392}},
  \bibinfo{pages}{232} (\bibinfo{year}{1983}).

\bibitem[{\citenamefont{Anghinolfi et~al.}(1996)\citenamefont{Anghinolfi,
  Ripani, Battaglieri, Cenni, Corvisiero et~al.}}]{Anghinolfi:1996vm}
\bibinfo{author}{\bibfnamefont{M.}~\bibnamefont{Anghinolfi}},
  \bibinfo{author}{\bibfnamefont{M.}~\bibnamefont{Ripani}},
  \bibinfo{author}{\bibfnamefont{M.}~\bibnamefont{Battaglieri}},
  \bibinfo{author}{\bibfnamefont{R.}~\bibnamefont{Cenni}},
  \bibinfo{author}{\bibfnamefont{P.}~\bibnamefont{Corvisiero}},
  \bibnamefont{et~al.}, \bibinfo{journal}{Nucl.Phys.}
  \textbf{\bibinfo{volume}{A602}}, \bibinfo{pages}{405} (\bibinfo{year}{1996}),
  \eprint{nucl-th/9603001}.

\bibitem[{\citenamefont{Anghinolfi et~al.}(1995)\citenamefont{Anghinolfi,
  Ripani, Cenni, Corvisiero, Longhi, Mazzaschi, Mokeev, Ricco, Taiuti, Teglia
  et~al.}}]{Anghinolfi:95n}
\bibinfo{author}{\bibfnamefont{M.}~\bibnamefont{Anghinolfi}},
  \bibinfo{author}{\bibfnamefont{M.}~\bibnamefont{Ripani}},
  \bibinfo{author}{\bibfnamefont{R.}~\bibnamefont{Cenni}},
  \bibinfo{author}{\bibfnamefont{P.}~\bibnamefont{Corvisiero}},
  \bibinfo{author}{\bibfnamefont{A.}~\bibnamefont{Longhi}},
  \bibinfo{author}{\bibfnamefont{L.}~\bibnamefont{Mazzaschi}},
  \bibinfo{author}{\bibfnamefont{V.}~\bibnamefont{Mokeev}},
  \bibinfo{author}{\bibfnamefont{G.}~\bibnamefont{Ricco}},
  \bibinfo{author}{\bibfnamefont{M.}~\bibnamefont{Taiuti}},
  \bibinfo{author}{\bibfnamefont{A.}~\bibnamefont{Teglia}},
  \bibnamefont{et~al.}, \bibinfo{journal}{Journal of Physics G: Nuclear and
  Particle Physics} \textbf{\bibinfo{volume}{21}}, \bibinfo{pages}{L9}
  (\bibinfo{year}{1995}).

\bibitem[{\citenamefont{Perdrisat et~al.}(2007)\citenamefont{Perdrisat,
  Punjabi, and Vanderhaeghen}}]{Perdrisat:2006hj}
\bibinfo{author}{\bibfnamefont{C.}~\bibnamefont{Perdrisat}},
  \bibinfo{author}{\bibfnamefont{V.}~\bibnamefont{Punjabi}}, \bibnamefont{and}
  \bibinfo{author}{\bibfnamefont{M.}~\bibnamefont{Vanderhaeghen}},
  \bibinfo{journal}{Prog. Part. Nucl. Phys.} \textbf{\bibinfo{volume}{59}},
  \bibinfo{pages}{694} (\bibinfo{year}{2007}).

\bibitem[{\citenamefont{Bradford et~al.}(2006)\citenamefont{Bradford, Bodek,
  Budd, and Arrington}}]{Bradford:2006yz}
\bibinfo{author}{\bibfnamefont{R.}~\bibnamefont{Bradford}},
  \bibinfo{author}{\bibfnamefont{A.}~\bibnamefont{Bodek}},
  \bibinfo{author}{\bibfnamefont{H.~S.} \bibnamefont{Budd}}, \bibnamefont{and}
  \bibinfo{author}{\bibfnamefont{J.}~\bibnamefont{Arrington}},
  \bibinfo{journal}{Nucl.Phys.Proc.Suppl.} \textbf{\bibinfo{volume}{159}},
  \bibinfo{pages}{127} (\bibinfo{year}{2006}), \eprint{hep-ex/0602017}.

\bibitem[{\citenamefont{Bernard et~al.}(2002)\citenamefont{Bernard,
  Elouadrhiri, and Meissner}}]{Bernard:2001rs}
\bibinfo{author}{\bibfnamefont{V.}~\bibnamefont{Bernard}},
  \bibinfo{author}{\bibfnamefont{L.}~\bibnamefont{Elouadrhiri}},
  \bibnamefont{and} \bibinfo{author}{\bibfnamefont{U.}~\bibnamefont{Meissner}},
  \bibinfo{journal}{J.Phys.} \textbf{\bibinfo{volume}{G28}},
  \bibinfo{pages}{R1} (\bibinfo{year}{2002}), \eprint{hep-ph/0107088}.

\bibitem[{\citenamefont{Sealock et~al.}(1989)\citenamefont{Sealock, Giovanetti,
  Thornton, Meziani, Rondon-Aramayo, Auffret, Chen, Christian, Day, McCarthy
  et~al.}}]{12C2}
\bibinfo{author}{\bibfnamefont{R.~M.} \bibnamefont{Sealock}},
  \bibinfo{author}{\bibfnamefont{K.~L.} \bibnamefont{Giovanetti}},
  \bibinfo{author}{\bibfnamefont{S.~T.} \bibnamefont{Thornton}},
  \bibinfo{author}{\bibfnamefont{Z.~E.} \bibnamefont{Meziani}},
  \bibinfo{author}{\bibfnamefont{O.~A.} \bibnamefont{Rondon-Aramayo}},
  \bibinfo{author}{\bibfnamefont{S.}~\bibnamefont{Auffret}},
  \bibinfo{author}{\bibfnamefont{J.-P.} \bibnamefont{Chen}},
  \bibinfo{author}{\bibfnamefont{D.~G.} \bibnamefont{Christian}},
  \bibinfo{author}{\bibfnamefont{D.~B.} \bibnamefont{Day}},
  \bibinfo{author}{\bibfnamefont{J.~S.} \bibnamefont{McCarthy}},
  \bibnamefont{et~al.}, \bibinfo{journal}{Phys. Rev. Lett.}
  \textbf{\bibinfo{volume}{62}}, \bibinfo{pages}{1350} (\bibinfo{year}{1989}).

\bibitem[{\citenamefont{Williamson et~al.}(1997)}]{Williamson:1997}
\bibinfo{author}{\bibfnamefont{C.}~\bibnamefont{Williamson}}
  \bibnamefont{et~al.}, \bibinfo{journal}{Phys. Rev.}
  \textbf{\bibinfo{volume}{C56}}, \bibinfo{pages}{3152} (\bibinfo{year}{1997}).

\bibitem[{\citenamefont{Andreopoulos et~al.}(2010)\citenamefont{Andreopoulos,
  Bell, Bhattacharya, Cavanna, Dobson et~al.}}]{Andreopoulos:2009rq}
\bibinfo{author}{\bibfnamefont{C.}~\bibnamefont{Andreopoulos}},
  \bibinfo{author}{\bibfnamefont{A.}~\bibnamefont{Bell}},
  \bibinfo{author}{\bibfnamefont{D.}~\bibnamefont{Bhattacharya}},
  \bibinfo{author}{\bibfnamefont{F.}~\bibnamefont{Cavanna}},
  \bibinfo{author}{\bibfnamefont{J.}~\bibnamefont{Dobson}},
  \bibnamefont{et~al.}, \bibinfo{journal}{Nucl.Instrum.Meth.}
  \textbf{\bibinfo{volume}{A614}}, \bibinfo{pages}{87} (\bibinfo{year}{2010}),
  \eprint{0905.2517}.

\bibitem[{\citenamefont{Dytman}(2011)}]{Dytman:2011zza}
\bibinfo{author}{\bibfnamefont{S.}~\bibnamefont{Dytman}}, \bibinfo{journal}{AIP
  Conf.Proc.} \textbf{\bibinfo{volume}{1382}}, \bibinfo{pages}{156}
  (\bibinfo{year}{2011}).

\bibitem[{\citenamefont{Dytman and Meyer}(2011)}]{Dytman:2014}
\bibinfo{author}{\bibfnamefont{S.~A.} \bibnamefont{Dytman}} \bibnamefont{and}
  \bibinfo{author}{\bibfnamefont{A.~S.} \bibnamefont{Meyer}},
  \bibinfo{journal}{AIP Conference Proceedings}
  \textbf{\bibinfo{volume}{1405}}, \bibinfo{pages}{213} (\bibinfo{year}{2011}).

\bibitem[{\citenamefont{O'Connell et~al.}(1987)\citenamefont{O'Connell, Dodge,
  Lightbody, Maruyama, Adler, Hansen, Schr\o{}der, Bernstein, Blomqvist,
  Cottman et~al.}}]{12C1}
\bibinfo{author}{\bibfnamefont{J.~S.} \bibnamefont{O'Connell}},
  \bibinfo{author}{\bibfnamefont{W.~R.} \bibnamefont{Dodge}},
  \bibinfo{author}{\bibfnamefont{J.~W.} \bibnamefont{Lightbody}},
  \bibinfo{author}{\bibfnamefont{X.~K.} \bibnamefont{Maruyama}},
  \bibinfo{author}{\bibfnamefont{J.~O.} \bibnamefont{Adler}},
  \bibinfo{author}{\bibfnamefont{K.}~\bibnamefont{Hansen}},
  \bibinfo{author}{\bibfnamefont{B.}~\bibnamefont{Schr\o{}der}},
  \bibinfo{author}{\bibfnamefont{A.~M.} \bibnamefont{Bernstein}},
  \bibinfo{author}{\bibfnamefont{K.~I.} \bibnamefont{Blomqvist}},
  \bibinfo{author}{\bibfnamefont{B.~H.} \bibnamefont{Cottman}},
  \bibnamefont{et~al.}, \bibinfo{journal}{Phys. Rev. C}
  \textbf{\bibinfo{volume}{35}}, \bibinfo{pages}{1063} (\bibinfo{year}{1987}).

\bibitem[{\citenamefont{Ankowski et~al.}(2010)\citenamefont{Ankowski, Benhar,
  and Farina}}]{Ankowski:2010yh}
\bibinfo{author}{\bibfnamefont{A.~M.} \bibnamefont{Ankowski}},
  \bibinfo{author}{\bibfnamefont{O.}~\bibnamefont{Benhar}}, \bibnamefont{and}
  \bibinfo{author}{\bibfnamefont{N.}~\bibnamefont{Farina}},
  \bibinfo{journal}{Phys.Rev.} \textbf{\bibinfo{volume}{D82}},
  \bibinfo{pages}{013002} (\bibinfo{year}{2010}), \eprint{1001.0481}.

\bibitem[{\citenamefont{Huber et~al.}(2009)\citenamefont{Huber, Lindner,
  Schwetz, and Winter}}]{Huber:2009cw}
\bibinfo{author}{\bibfnamefont{P.}~\bibnamefont{Huber}},
  \bibinfo{author}{\bibfnamefont{M.}~\bibnamefont{Lindner}},
  \bibinfo{author}{\bibfnamefont{T.}~\bibnamefont{Schwetz}}, \bibnamefont{and}
  \bibinfo{author}{\bibfnamefont{W.}~\bibnamefont{Winter}},
  \bibinfo{journal}{JHEP} \textbf{\bibinfo{volume}{0911}}, \bibinfo{pages}{044}
  (\bibinfo{year}{2009}), \eprint{0907.1896}.

\bibitem[{\citenamefont{Coloma and Huber}(2013)}]{Coloma:2013rqa}
\bibinfo{author}{\bibfnamefont{P.}~\bibnamefont{Coloma}} \bibnamefont{and}
  \bibinfo{author}{\bibfnamefont{P.}~\bibnamefont{Huber}},
  \bibinfo{journal}{Phys.Rev.Lett.} \textbf{\bibinfo{volume}{111}},
  \bibinfo{pages}{221802} (\bibinfo{year}{2013}), \eprint{1307.1243}.

\bibitem[{\citenamefont{Coloma et~al.}(2014)\citenamefont{Coloma, Huber, Jen,
  and Mariani}}]{Coloma:2013tba}
\bibinfo{author}{\bibfnamefont{P.}~\bibnamefont{Coloma}},
  \bibinfo{author}{\bibfnamefont{P.}~\bibnamefont{Huber}},
  \bibinfo{author}{\bibfnamefont{C.-M.} \bibnamefont{Jen}}, \bibnamefont{and}
  \bibinfo{author}{\bibfnamefont{C.}~\bibnamefont{Mariani}},
  \bibinfo{journal}{Phys.Rev.} \textbf{\bibinfo{volume}{D89}},
  \bibinfo{pages}{073015} (\bibinfo{year}{2014}), \eprint{1311.4506}.

\bibitem[{\citenamefont{Huber et~al.}(2005)\citenamefont{Huber, Lindner, and
  Winter}}]{globes1}
\bibinfo{author}{\bibfnamefont{P.}~\bibnamefont{Huber}},
  \bibinfo{author}{\bibfnamefont{M.}~\bibnamefont{Lindner}}, \bibnamefont{and}
  \bibinfo{author}{\bibfnamefont{W.}~\bibnamefont{Winter}},
  \bibinfo{journal}{Comput.Phys.Commun.} \textbf{\bibinfo{volume}{167}},
  \bibinfo{pages}{195} (\bibinfo{year}{2005}), \eprint{hep-ph/0407333}.

\bibitem[{\citenamefont{Huber et~al.}(2007)\citenamefont{Huber, Kopp, Lindner,
  Rolinec, and Winter}}]{globes2}
\bibinfo{author}{\bibfnamefont{P.}~\bibnamefont{Huber}},
  \bibinfo{author}{\bibfnamefont{J.}~\bibnamefont{Kopp}},
  \bibinfo{author}{\bibfnamefont{M.}~\bibnamefont{Lindner}},
  \bibinfo{author}{\bibfnamefont{M.}~\bibnamefont{Rolinec}}, \bibnamefont{and}
  \bibinfo{author}{\bibfnamefont{W.}~\bibnamefont{Winter}},
  \bibinfo{journal}{Comput.Phys.Commun.} \textbf{\bibinfo{volume}{177}},
  \bibinfo{pages}{432} (\bibinfo{year}{2007}), \eprint{hep-ph/0701187}.

\end{thebibliography}
%
\end{document}